\documentclass[twocolumn,apj,floatfix]{emulateapj}

\usepackage{natbib}
\newcommand{\beq}{\begin{equation}}
\newcommand{\eeq}{\end{equation}}
\newcommand{\bdm}{\begin{displaymath}}
\newcommand{\edm}{\end{displaymath}}
\newcommand{\bea}{\begin{eqnarray}}
\newcommand{\eea}{\end{eqnarray}}
\newcommand{\bt}{\begin{tabular}}
\newcommand{\et}{\end{tabular}}
\newcommand{\xv}{{\bf x}}
\newcommand{\kv}{{\bf k}}

\newcommand{\qv}{{\bf q}}

\newcommand{\de}{{\rm d}}
\newcommand{\ep}{\epsilon}

\def\d{\delta}
\def\De{\Delta}

\def\cGpc{\, h^{-3} \, {\rm Gpc}^3}
\def\kMpc{\, h \, {\rm Mpc}^{-1}}
\def\icMpc{\, h^3 \, {\rm Mpc}^{-3}}
\def\fNL{f_{NL}}

\def\fNLl{f_{NL}^{\rm loc.}}
\def\dfNLl{\Delta f_{NL}^{\rm loc.}}
\def\fNLe{f_{NL}^{\rm eq.}}
\def\dfNLe{\Delta f_{NL}^{\rm eq.}}
\def\kMAX{k_{\rm max}}
\def\pd{\partial}
\def\dbo{\Delta b_1}
\def\dbt{\Delta b_2}

\begin{document}

\preprint{FERMILAB-PUB-00-000-0}

\title{The bispectrum of galaxies from high-redshift galaxy surveys:\\
primordial non-Gaussianity and non-linear galaxy bias}

  \author{Emiliano Sefusatti\altaffilmark{1}}
  \email{emiliano@fnal.gov}
  \altaffiltext{1}{Particle Astrophysics Center, Fermi National Accelerator 
Laboratory, Batavia, IL 60510-0500}
  \author{Eiichiro Komatsu\altaffilmark{2}}
  \altaffiltext{2}{Department of Astronomy, University of Texas at Austin, 
2511 Speedway, RLM 15.306, TX 78712}

\begin{abstract}

The greatest challenge in the interpretation of galaxy clustering data from
 any surveys is galaxy bias. Using a simple Fisher matrix
 analysis, we show that the bispectrum
 provides an excellent determination of linear and non-linear bias
 parameters of intermediate and high-$z$ galaxies, when all measurable triangle
 configurations down to mildly non-linear scales, where perturbation
 theory is still valid, are included. The bispectrum is also a powerful
 probe of primordial non-Gaussianity. The
 planned galaxy surveys at $z\gtrsim 2$ should yield constraints on
 non-Gaussian parameters, $\fNLl$ and $\fNLe$, that are comparable to,
 or even better than, those from CMB experiments. We study how these
 constraints improve with volume, redshift range, as well as the number
 density of galaxies. Finally, we show that a halo occupation
 distribution may be used to improve these
 constraints further by lifting degeneracies between gravity, bias, and
 primordial non-Gaussianity.

\end{abstract}

\keywords{cosmology: theory - large-scale structure of the Universe}

\maketitle

\section{Introduction}

Why study high-$z$ galaxy surveys? 
The recognition that baryon acoustic oscillations in the galaxy power
spectrum \citep{ColeEtal2005,EisensteinEtal2005,Hutsi2006,PercivalEtal2007}
are an excellent probe of the nature of dark energy has led to several 
 proposals for large-volume redshift surveys at $z\gtrsim 1$.

Galaxies are a biased tracer of the underlying matter distribution.
The use of highly biased tracers, such as luminous red galaxies at lower
$z$ and Lyman break galaxies or Lyman-$\alpha$ emitters at 
higher $z$ requires a reliable modelling of non-linearity and 
scale-dependence of galaxy bias, even at relatively large spatial scales
 \citep{SmithScoccimarroSheth2006,SmithScoccimarroSheth2007,McDonald2006}.

The distribution of galaxies is non-Gaussian. The galaxy bispectrum, the 
three-point correlation function in Fourier space, does not vanish. 
It has been known for more than a decade that the bispectrum is an
excellent tool for measuring galaxy bias parameters, independent of the 
overall 
normalization of dark matter fluctuations \citep{Fry1994,
MatarreseVerdeHeavens1997,ScoccimarroEtal2001B}. 
This method has been applied successfully to existing galaxy surveys such as 
the 2dFGRS and the SDSS, yielding constraints on non-linearities in galaxy 
bias \citep{VerdeEtal2002,PanSzapudi2005,GaztanagaEtal2005,NishimichiEtal2006} 
as well as on the Halo Occupation Distribution (HOD) \citep{KulkarniEtal2007}. 
Moreover, it has been shown that the galaxy bispectrum contains additional 
cosmological information that is not present in the power spectrum 
\citep{SefusattiScoccimarro2005,SefusattiEtal2006}.

The galaxy bispectrum on large scales, or other statistical tools that
are sensitive to the higher-order correlation of galaxies, are sensitive to 
statistical properties of primordial fluctuations: primordial non-Gaussianity 
\citep{FryScherrer1994,ChodorowskiBouchet1996,Scoccimarro2000A,VerdeEtal2000,
DurrerEtal2000,ScoccimarroSefusattiZaldarriaga2004,
HikageKomatsuMatsubara2006}. The common belief is that the cosmic microwave 
background (CMB) is most sensitive to primordial non-Gaussianity; however, 
the number of Fourier modes available in the galaxy survey data is much larger 
than that in the CMB data, as the former probes the full three-dimensional
structure of density fields. The galaxy bispectrum can, therefore,
outperform the CMB bispectrum, provided that the other sources of
non-Gaussianity, such as non-linear gravitational evolution and galaxy
bias, are under control.
Furthermore, it should always be emphasized that galaxy surveys provide
information on the spatial scales that are much smaller than those
probed by CMB; thus, these two measurements are complementary to each other.

Motivated by these considerations, in this paper we study how well the planned 
high-$z$ galaxy surveys would constrain primordial non-Gaussianity and galaxy 
bias parameters. High-$z$ surveys are more useful for this task than low-$z$ 
ones owing to much weaker non-linearities in matter clustering and 
redshift-space distortion, which allows us to use the galaxy bispectrum to the 
smaller spatial scales that are inaccessible by low-$z$ surveys.

This paper is organized as follows. In section \ref{sec_bispectra} we describe 
the leading contributions to the bispectrum of the observed galaxy 
distribution. In section \ref{sec_fisher} we present our Fisher matrix 
analysis for the galaxy bispectrum, and in section \ref{sec_results} we show 
our predictions for constraints on galaxy bias and primordial non-Gaussianity 
from sample galaxy redshift survey designs, studying dependence on the survey 
volume, maximum wavenumber, number density of galaxies, and redshifts. We also 
present our predictions for the ongoing, upcoming and planned galaxy surveys. 
In section \ref{sec_hod} we extend our fisher matrix analysis to the halo 
occupation distribution as a tool to describe galaxy biasing at large scales. 
Finally, in section \ref{sec_conclusions} we present our conclusions.

\section{The Galaxy Bispectrum}
\label{sec_bispectra}

\subsection{Primordial non-Gaussianity}
\label{subsec_shapes}

We explore two parametrizations of primordial non-Gaussianity which are 
motivated by inflationary models. While representing a wide variety of 
non-Gaussian models, these parametrizations are by no means exhaustive. Our 
method can be applied to any other functional forms of the bispectrum 
\citep[e.g.,][]{LiguoriEtal2006,ChenEtal2006} in a straightforward way.

\subsubsection{Local model}

The first one is described by the {\it local} expression for Bardeen's
curvature perturbations during the matter era, $\Phi(x)$, in position
space
\citep{GanguiEtal1994,VerdeEtal2000,KomatsuSpergel2001}
\beq
\Phi(\xv)=\Phi_G(\xv)+\fNLl[\Phi_G^2(\xv)-\langle\Phi_G^2(\xv)\rangle],
\label{deffNL}
\eeq
where $\Phi_G(\xv)$ is a Gaussian field and $\fNLl$ 
is a constant characterizing the amplitude of primordial
non-Gaussianity. In this case, the leading contribution in the $\fNLl$
expansion  
to the bispectrum, $B_{\Phi}^{local}(k_1,k_2,k_3)$, of the curvature field 
is given by 
\bea\label{BispPrimLocal}
B_{\Phi}^{local} & \simeq & 2\fNLl[P_{\Phi}(k_1)P_{\Phi}(k_2)+{\rm cyc.}]
\nonumber\\
& = & 2\fNLl C_\Phi^2\left[\frac{1}{k_1^{4-n_s} k_2^{4-n_s}}+{\rm cyc.}\right],
\eea
where we approximate $P_{\Phi}(k)\simeq P_{\Phi_G}(k)$, and 
\beq
C_\Phi\equiv \frac{P_\Phi(k)}{k^{n_s-4}},
\eeq
which quantifies departure from a scale-invariant spectrum.
We include  $n_s$ explicitly, as we are interested in determining how 
a departure from scale invariance, which has been hinted by WMAP 
\citep{SpergelEtal2006}, would affect detectability of primordial
non-Gaussianity in the distribution of galaxies. For this local model, most of 
the signal is given by squeezed triangular configurations, $k_1\ll k_2$, $k_3$.
The {\it local} type of non-Gaussianity described by equation
(\ref{BispPrimLocal}) 
is predicted in models such as the curvaton scenario \citep{LythEtal2003}, 
models with inhomogeneous reheating 
\citep{DvaliGruzinovZaldarriaga2004A,DvaliGruzinovZaldarriaga2004B}, multiple 
field inflationary models \citep{BernardeauUzan2002} or generically in models
where the non-linearities arise from the evolution of perturbations outside 
the horizon. 

The best limits to date on possible values for the $\fNLl$ parameter come from 
measurements of the microwave background bispectrum on the WMAP data
\citep{KomatsuEtal2003,SpergelEtal2006,CreminelliEtal2007} $-36\leq \fNLl 
\leq 100$ at 95\% C.L., which corresponds to the 1-$\sigma$ error of
\beq
\dfNLl = 34\quad ({\rm WMAP3}),
\eeq
which is a factor of 50 better than the limit from COBE 
\citep{KomatsuEtal2002}. Upon completion, WMAP is expected to reach 
$\dfNLl\simeq 20$, while the Planck satellite would yield $\dfNLl\simeq 3$ 
\citep{KomatsuSpergel2001,BabichZaldarriaga2004,YadavEtal2007}. 

Measurements of the galaxy bispectrum in the SDSS main sample are expected to 
yield $\dfNLl\simeq 150$ \citep{ScoccimarroSefusattiZaldarriaga2004}. Recently 
\citet{PillepichEtal2006} pointed out that a full-sky measurement of the 
bispectrum of fluctuations in the 21-cm background might reach 
$\fNLl\simeq 1$, while a more aggressive analysis by \citet{Cooray2006} shows 
that the same observations could reach $\dfNLl\simeq 0.01$ in principle. Other 
large-scale structure probes such as cluster abundance, on the other hand, is 
unlikely to improve CMB limits on $\fNLl$ \citep{SefusattiEtal2007}; 
however,  it should provide an important cross-check of the results if a 
significant $\fNLl$ was detected in the CMB, and it should not be forgotten 
that the spatial scales probed by the cluster abundance is smaller than those 
probed by the CMB.

\subsubsection{Equilateral model}

The second model for primordial non-Gaussianity is given by
\bea
B_{\Phi}^{equil.} & = & 6\fNLe C_\Phi^2\left[
-\frac{1}{k_1^{4-n_s} k_2^{4-n_s}}-\frac{1}{k_1^{4-n_s} k_3^{4-n_s}}
\right.
\nonumber\\
& & -\frac{1}{k_2^{4-n_s} k_2^{4-n_s}}-\frac{2}{(k_1k_2k_3)^{2(4-n_s)/3}}
\nonumber\\
& & \left.
+\left(\frac{1}{k_1^{(4-n_s)/3}k_2^{2(4-n_s)/3}k_3^{4-n_s}}
+{\rm cyc.}\right)\right].
\label{BispPrimEqui}
\eea
 \citet{BabichCreminelliZaldarriaga2004} and 
\citet{CreminelliEtal2007} have shown that this form provides a good
approximation to the bispectra predicted by 
higher derivatives and DBI inflationary models \citep{Creminelli2003,
AlishahihaEtal2004}.

The bispectrum in equation (\ref{BispPrimEqui}) is normalized in a such a way 
that for equilateral configurations ($k_1=k_2=k_3=k$), it coincides with the 
local form given in equation (\ref{BispPrimLocal}). 
The important difference is that this form has the largest contribution
from the equilateral configurations, as opposed to the local form in which
the largest contribution comes from the squeezed configurations.
The current limits from WMAP are $-256\leq \fNLe \leq 332$ 
at 95\% C.L. \citep{CreminelliEtal2007}, which corresponds to the
1-$\sigma$ error of
\beq
\dfNLe = 147\quad ({\rm WMAP3}).
\eeq

\subsubsection{The primordial density bispectrum}

Density fluctuations in Fourier space, $\delta_{\mathbf k}$, are related to 
the curvature perturbations, $\Phi_{\mathbf k}$, via the Poisson equation,
$\delta_{\mathbf k}(a)=M(k;a)\Phi_{\mathbf k}$, where 
\beq
M(k,a)\equiv \frac{2}{3}\frac{D(a)}{H_o^2\Omega_m}k^2 T(k).
\label{eq:mk}
\eeq
Here $a$ is the scale factor, $T(k)$ is the matter transfer function, and 
$D(a)$ is the growth function
\footnote{The function, $M(k;a)$, uses the same definition as in 
 \citet{VerdeEtal2000} for $\Omega_m=1$, but it differs from $M(k)$ given in 
\citet{HikageKomatsuMatsubara2006}, where the dependence on the growth 
function has been taken out as $M(k;a)=M^{HKM}(k)D(a)$. Also, the same 
function is defined in \citet{ScoccimarroSefusattiZaldarriaga2004} in terms of 
the gravitational potential (i.e., the 0-0 component of the metric 
perturbations) during radiation domination, so that 
$M(k;a)=-\frac{9}{10}M^{SSZ}(k;a)$ leading to a different definition of the 
non-Gaussian parameter, $\fNL=-\frac{10}{9}\fNL^{SSZ}$}.

This allows us to write the primordial contribution to a generic $n$-point 
function of the matter density fields in terms of the respective
correlator of the curvature perturbations as
\bea\label{npoint}
\langle\d_{\kv_1}\d_{\kv_2}... \d_{\kv_N}\rangle_I & = &
M(k_1;a)M(k_2;a)... M(k_N;a)\nonumber\\
& & \times
\langle\Phi_{\kv_1}\Phi_{\kv_2}... \Phi_{\kv_N}\rangle.
\eea
In particular, the initial (primordial) matter bispectrum, $B_I(k_1,k_2,k_3)$, 
is given by 
\bea
\label{eq:bi}
B_I(k_1,k_2,k_3) & = & M(k_1)M(k_2)M(k_3)
\nonumber \\
& & \times B_{\Phi}(k_1,k_2,k_3),
\eea
where we have omitted for brevity the explicit dependence of $M(k;a)$ on $a$.
We can also relate the linear density power spectrum, $P_L(k)$, to the 
curvature power spectrum, $P_{\Phi}(k)$, as
\beq
P_L(k)=M^2(k)P_{\Phi}(k).
\eeq

A hierarchical relation between the scale dependence of the initial bispectrum 
and the power spectrum such as $B_\Phi(k)\sim \fNL P_{\Phi}^2(k)$ with a 
constant $\fNL$ is by no means generic or universal. A $\fNL$ with a peculiar 
scale dependence unrelated to the power spectrum appears, for instance, in 
string-motivated models such as DBI inflation \citep{AlishahihaEtal2004,
Chen2005}. Our analysis can be applied to any models of primordial
non-Gaussianity, provided that the bispectrum can be calculated from those 
models. 

Note that a post-Newtonian effect can yield an additional contribution to 
non-linearity of primordial perturbations and hence to non-Gaussianity 
\citep{BartoloMatarreseRiotto2005}. Although we do not include this effect our 
analysis, it would be interesting to study how important the post-Newtonian 
effect would be for the future galaxy surveys.

\subsection{Non-Gaussianity from non-linear gravitational evolution}

Even if the initial perturbations are Gaussian, the subsequent gravitational 
evolution makes the evolved density fields non-Gaussian. On large scales one 
can study the non-linear evolution of matter density fluctuations by means of 
perturbation theory, and write the solution up to the second order in  $\d$ as
\beq
\d_{\kv} \simeq  \d^{(1)}_{\kv}\!+\!\!\!\int\!\!\!\de^3q_1\de^3q_2
\d_D(\kv\! -\!\qv_{12})
F_2(\qv_1,\qv_2)\d^{(1)}_{\qv_1}\d^{(1)}_{\qv_2},
\label{deltaNL}
\eeq
where $\d^{(1)}$ is the linear solution, and
$F_2(\kv_1,\kv_2)$ is a known mathematical function given by
\beq
F_2(\kv_1,\kv_2) = \frac{5}{7}+\frac{x}{2}\left(\frac{k_1}{k_2}
+\frac{k_2}{k_1}\right)+\frac{2}{7}\ x^2,
\eeq
with $x\equiv\hat{\kv}_1\cdot\hat{\kv_2}$.
Therefore, one obtains
\beq 
B_G(k_1,k_2,k_3)= 2F_2(\kv_1,\kv_2)P_L(k_1)P_L(k_2)+{\rm cyc.}
\label{eq:bg}
\eeq

\begin{figure*}[t]
\begin{center}
\includegraphics[width=0.45\textwidth]{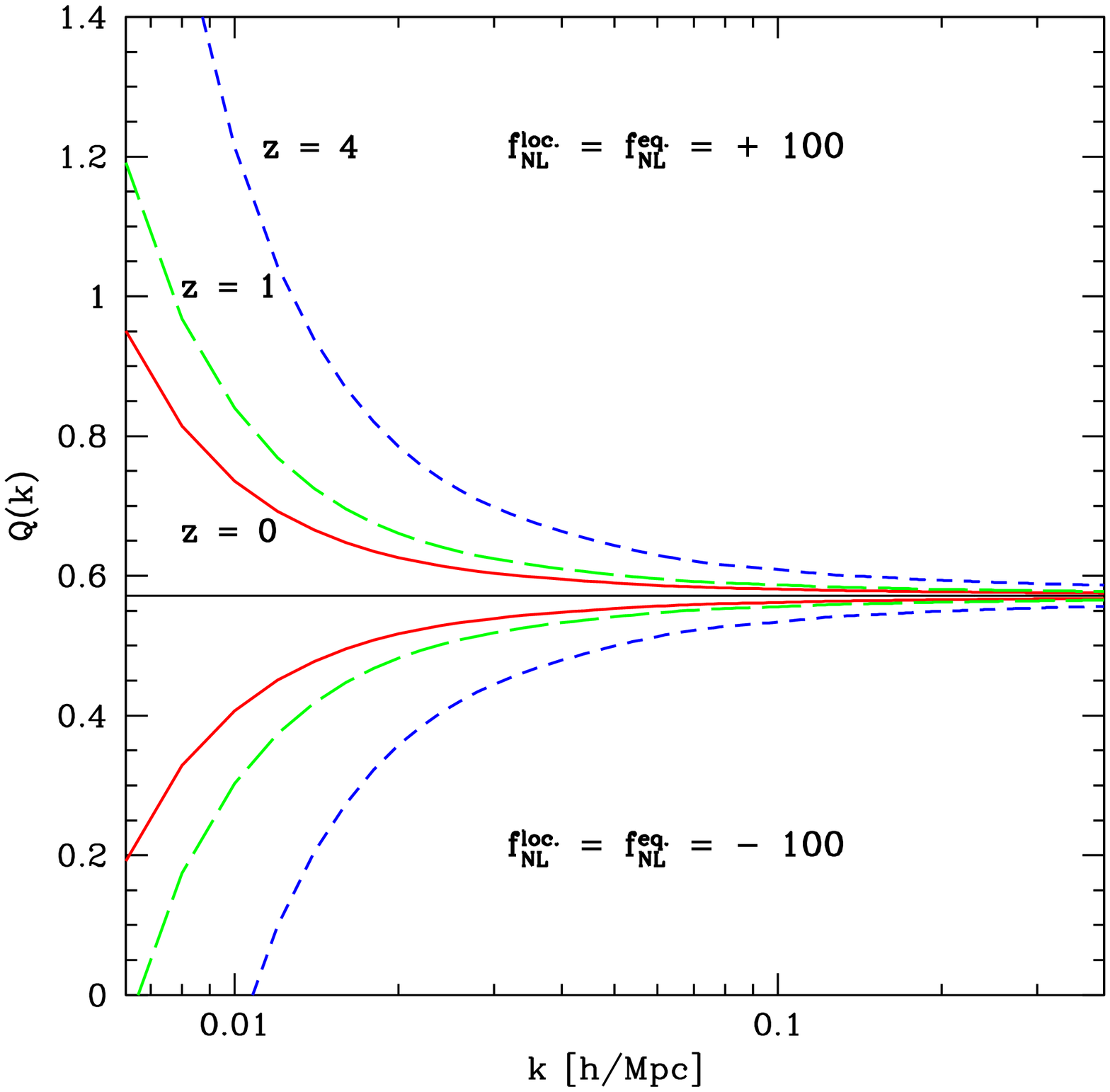}
\includegraphics[width=0.45\textwidth]{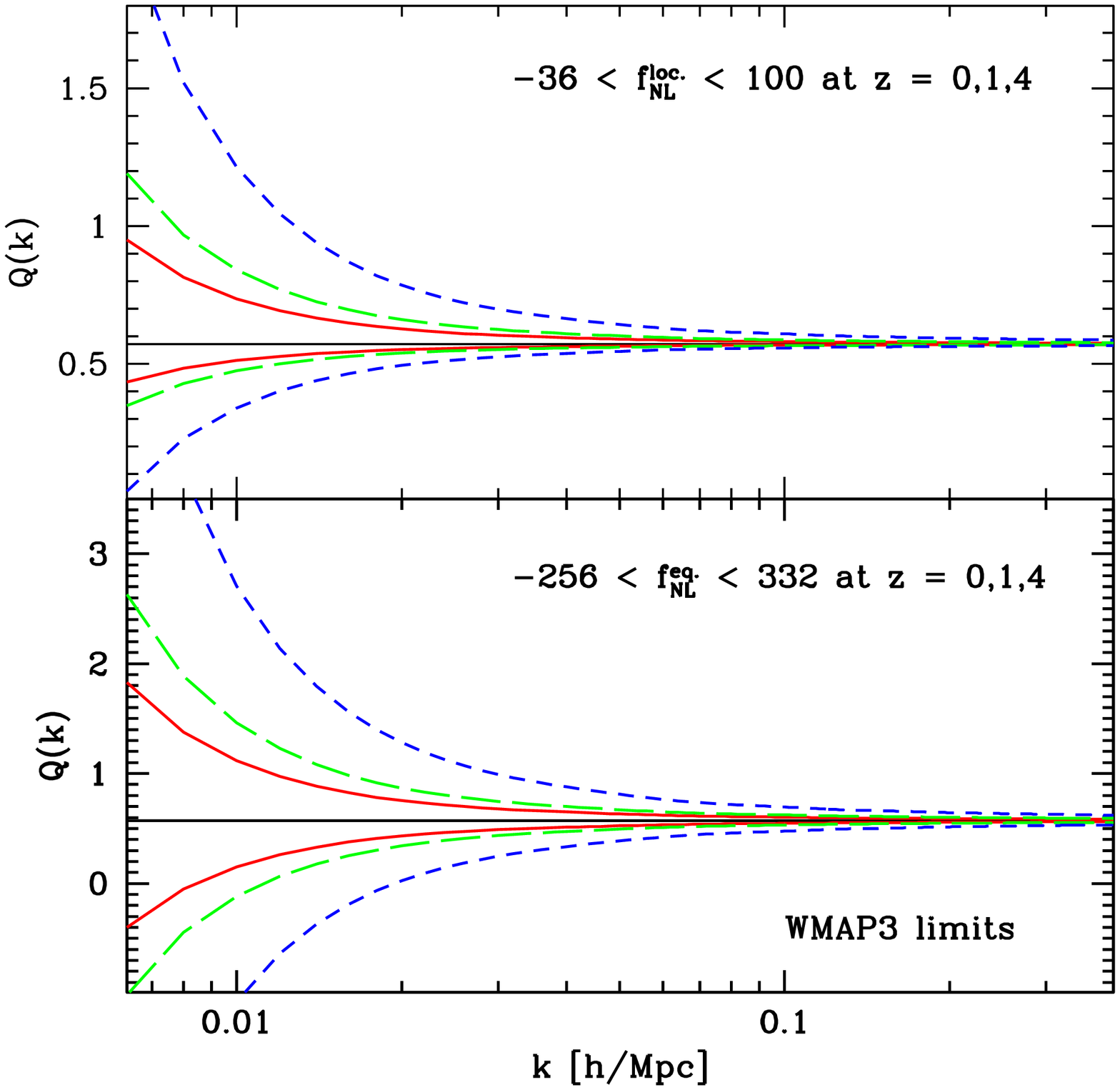}
\caption{\label{Fig_NGShapes_eq} 
Equilateral configurations of the reduced bispectrum of dark matter
 distribution at the second order (``tree-level'').
The horizontal lines at $Q(k)=0.57$ show the gravitational contribution
 only, which corresponds to $\fNLl=0=\fNLe$. The solid, long-dashed and
 short-dashed lines show the gravitational contribution with
 non-Gaussian initial perturbations at $z=0$, 1, and 4,
 respectively. Note that  $\fNLl=\fNLe$ for equilateral configurations.
({\it Left panel}) The curves above $Q(k)=0.57$ show $f_{NL}=+100$,
while the curves below it show $f_{NL}=-100$.
({\it Right panel}) The same as the left panel but for the current 
WMAP3 limits on $\fNLl$ (top) and $\fNLe$ (bottom).}
\end{center}
\end{figure*}

The bispectrum of matter density fluctuations (i.e., no galaxies yet) evolved 
from non-Gaussian primordial fluctuations on large scales is thus given by the 
sum of equation~(\ref{eq:bi}) and (\ref{eq:bg}):
\beq
B(k_1,k_2,k_3)=B_I(k_1,k_2,k_3)+B_G(k_1,k_2,k_3). 
\eeq

As usual, we shall focus on the {\it reduced} bispectrum, defined as
\beq
Q(k_1,k_2,k_3)\equiv\frac{B(k_1,k_2,k_3)}{P(k_1)P(k_2)+{\rm cyc.}}
\eeq
which has an advantage of being only mildly sensitive to cosmological 
parameters. That is to say, the dependence on cosmology has been ``factored 
out'' by a product of the power spectra in the denominator and the $Q_G$ 
component is particularly insensitive to the amplitude of matter fluctuations 
(e.g., $\sigma_8$). The reduced bispectrum of matter density fluctuations is 
also given by the sum of two contributions:
\bea
Q(k_1,k_2,k_3) & = & Q_I(k_1,k_2,k_3)+Q_G(k_1,k_2,k_3)
\nonumber\\
& =& \frac{B_I(k_1,k_2,k_3)}{P(k_1)P(k_2)+{\rm cyc.}}\nonumber\\
& &  +\frac{B_G(k_1,k_2,k_3)}{P(k_1)P(k_2)+{\rm cyc.}}
\eea
It is important to remember that, in the leading order, 
$Q_G$ does not depend on the linear growth factor, $D(a)$, and thus it
is independent of redshifts. In other words, $B_G$ is proportional to
$D^4$, which cancels $D^4$ in $[P(k)]^2$ in the denominator exactly.
On the other hand, $Q_I$ is proportional to $1/D(a)$ because $B_I\propto
D^3$, and thus it is
larger at higher redshifts. Therefore, high-$z$ galaxy surveys are expected to 
be more sensitive to primordial non-Gaussianity, relative to the
gravitational bispectrum, than low-$z$ ones. 

In Figure \ref{Fig_NGShapes_eq} we plot the equilateral configurations
of the reduced bispectrum, $Q(k)\equiv Q(k,k,k)$, from
non-linear gravitational evolution and non-Gaussian initial conditions 
at $z=0$, 1, and 4. 
As mentioned earlier the local and equilateral model of primordial
non-Gaussianity give the same results for these configurations. 
For Gaussian initial fluctuations, $f_{NL}=0$, $Q(k)=0.57$ at tree-level in 
perturbation theory and is independent of scales.\footnote{The equilateral 
reduced bispectrum is independent of scales only in the second order 
perturbations. A scale dependence arises when the higher-order terms are 
included \citep{ScoccimarroEtal1998}.} On the other hand, $Q(k)$ exhibits a 
clear scale dependence for $f_{NL}\neq 0$. A positive $f_{NL}$ enhances $Q(k)$ 
at large scales, i.e., $Q(k)>0.57$, whereas a negative $f_{NL}$ suppresses it. 
This is because a positive $f_{NL}$ results in positively skewed density
fluctuations. It is easy to show that the scale dependence of the primordial 
component $Q_I(k)$ is given by $1/M(k)$, where $M(k)$ is given by 
equation~(\ref{eq:mk}).
As $1/M(k)\propto k^{-2}$ on large scales, we find
that the primordial non-Gaussian signal is larger on 
large scales. This property makes it easier to find primordial
non-Gaussianity in CMB observations; however, 
as there are much more modes available on smaller scales, 
 the {\it cumulative} signal to noise for higher-order correlation functions 
increases due to the large number of observable configurations on small scales
\citep{SefusattiScoccimarro2005}. 

\begin{figure*}[t]
\begin{center}
\bt{cc}
\includegraphics[width=0.45\textwidth]{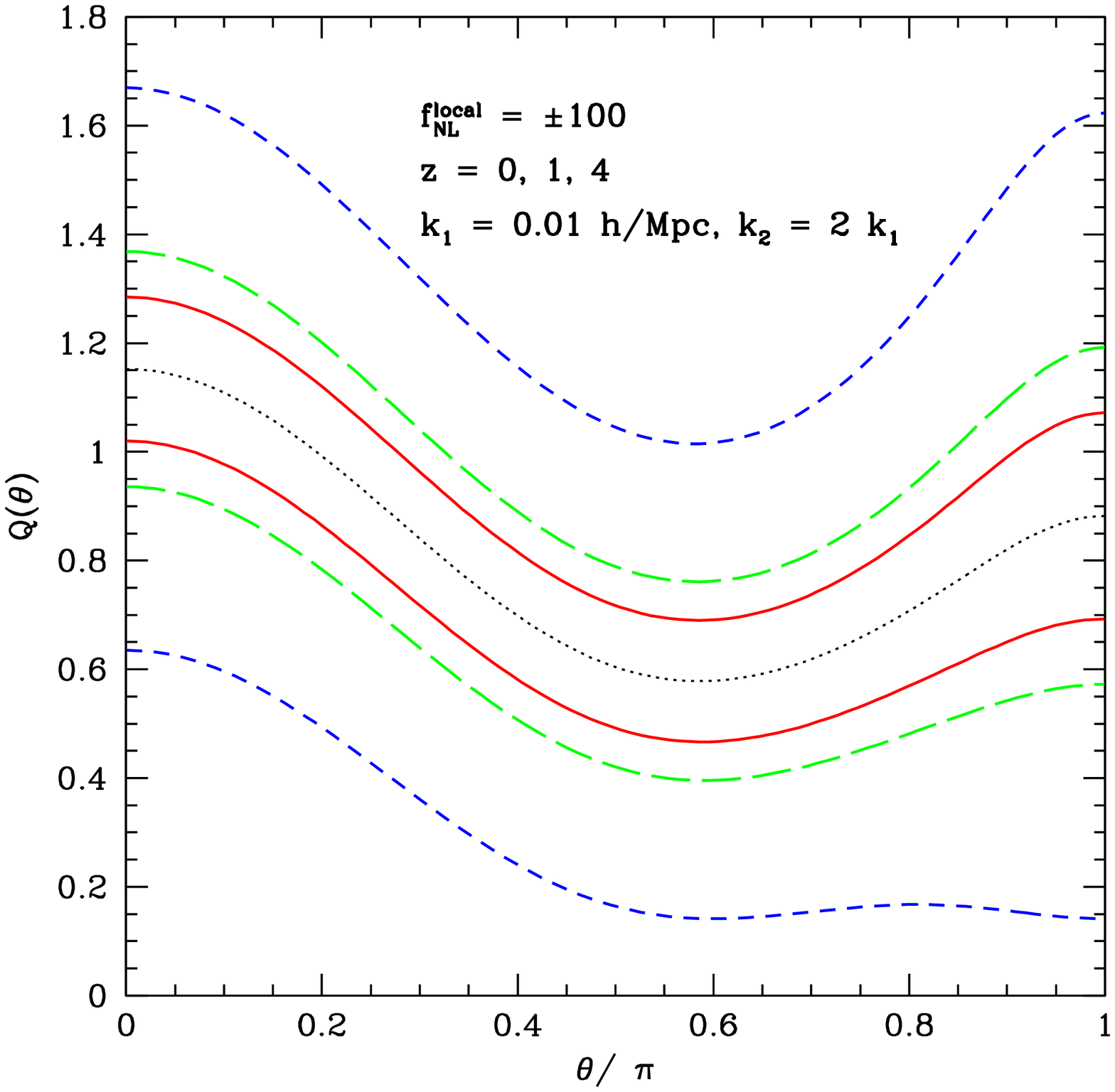} &
\includegraphics[width=0.45\textwidth]{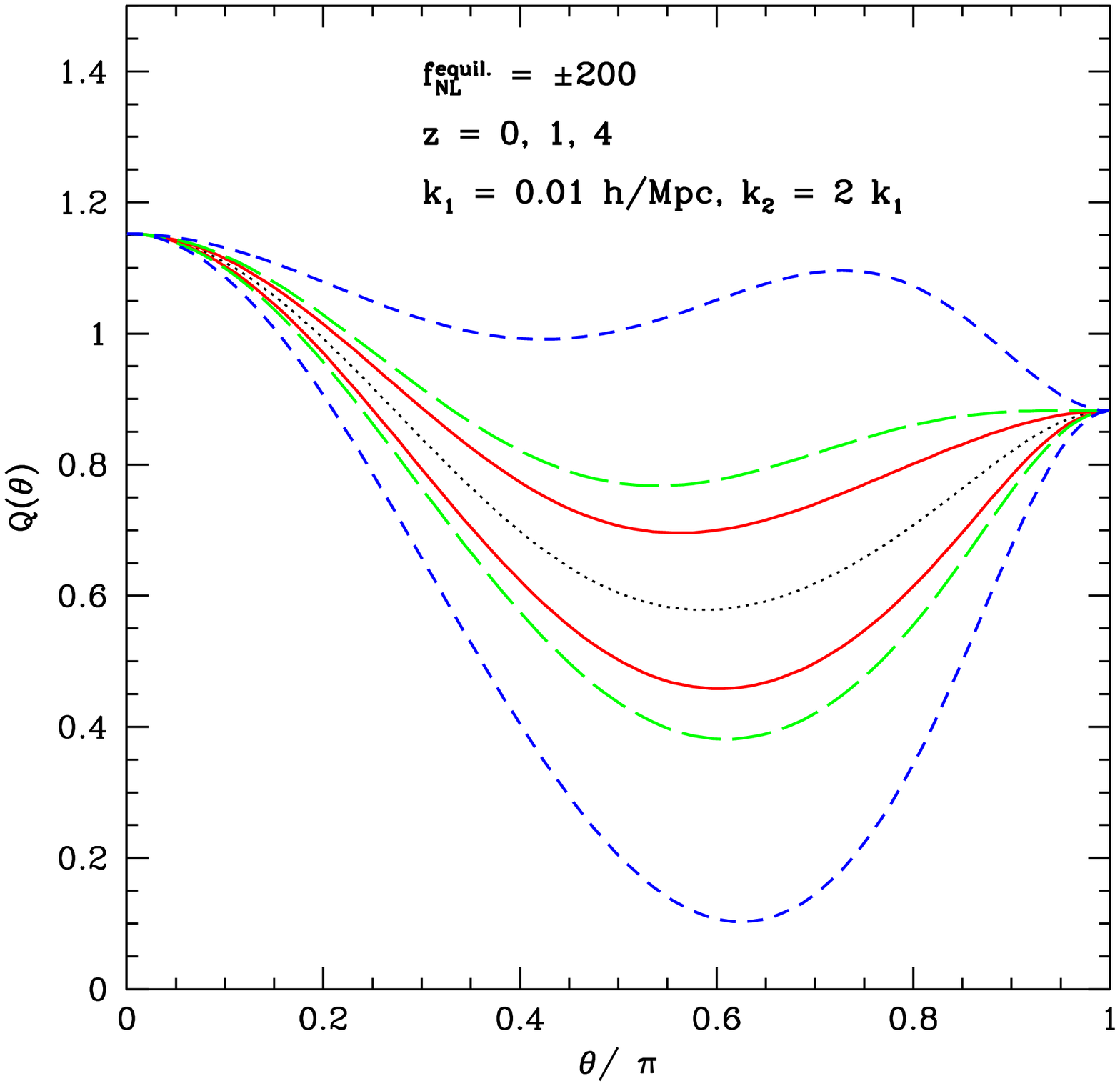} \\
\includegraphics[width=0.45\textwidth]{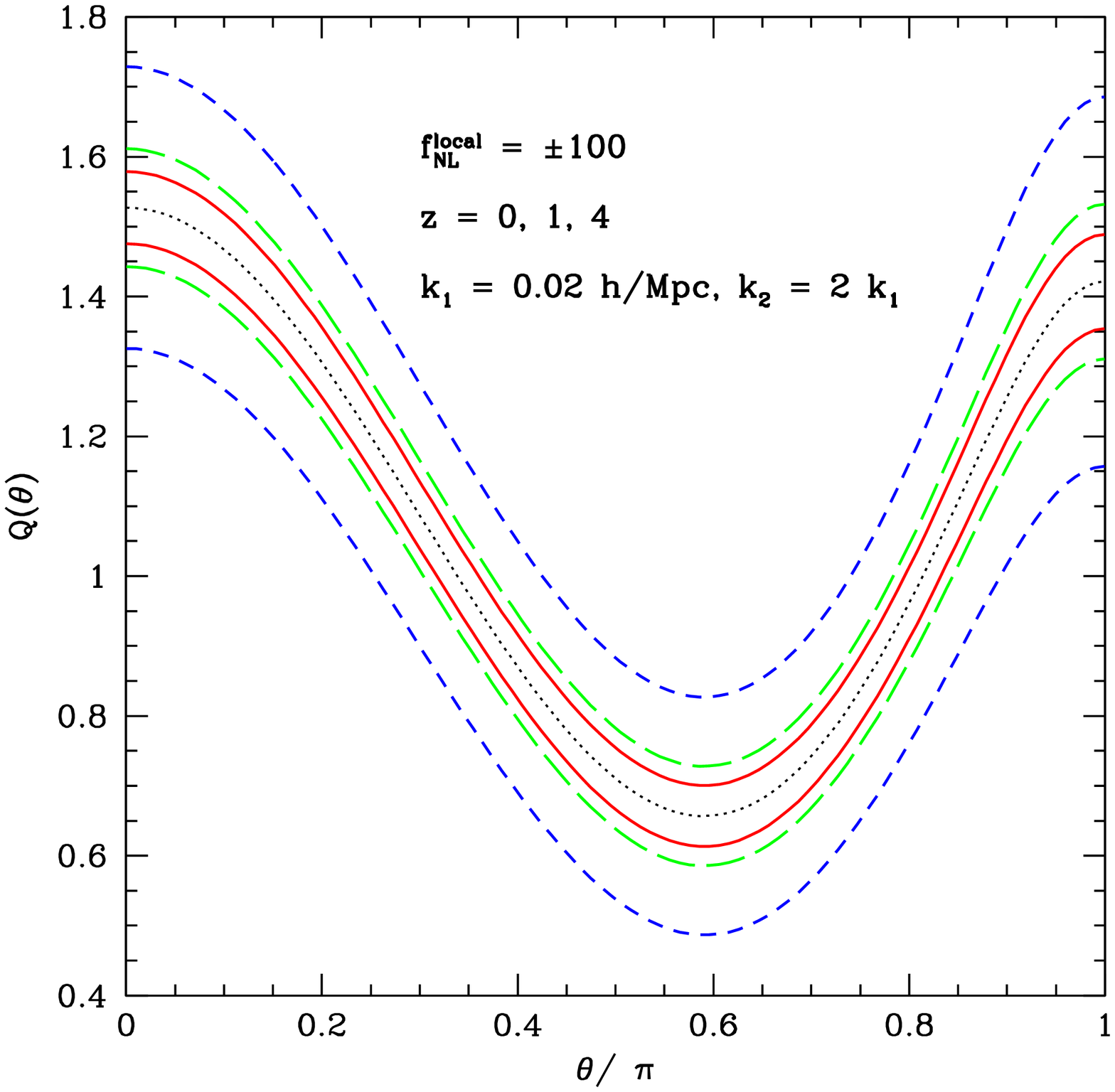} &
\includegraphics[width=0.45\textwidth]{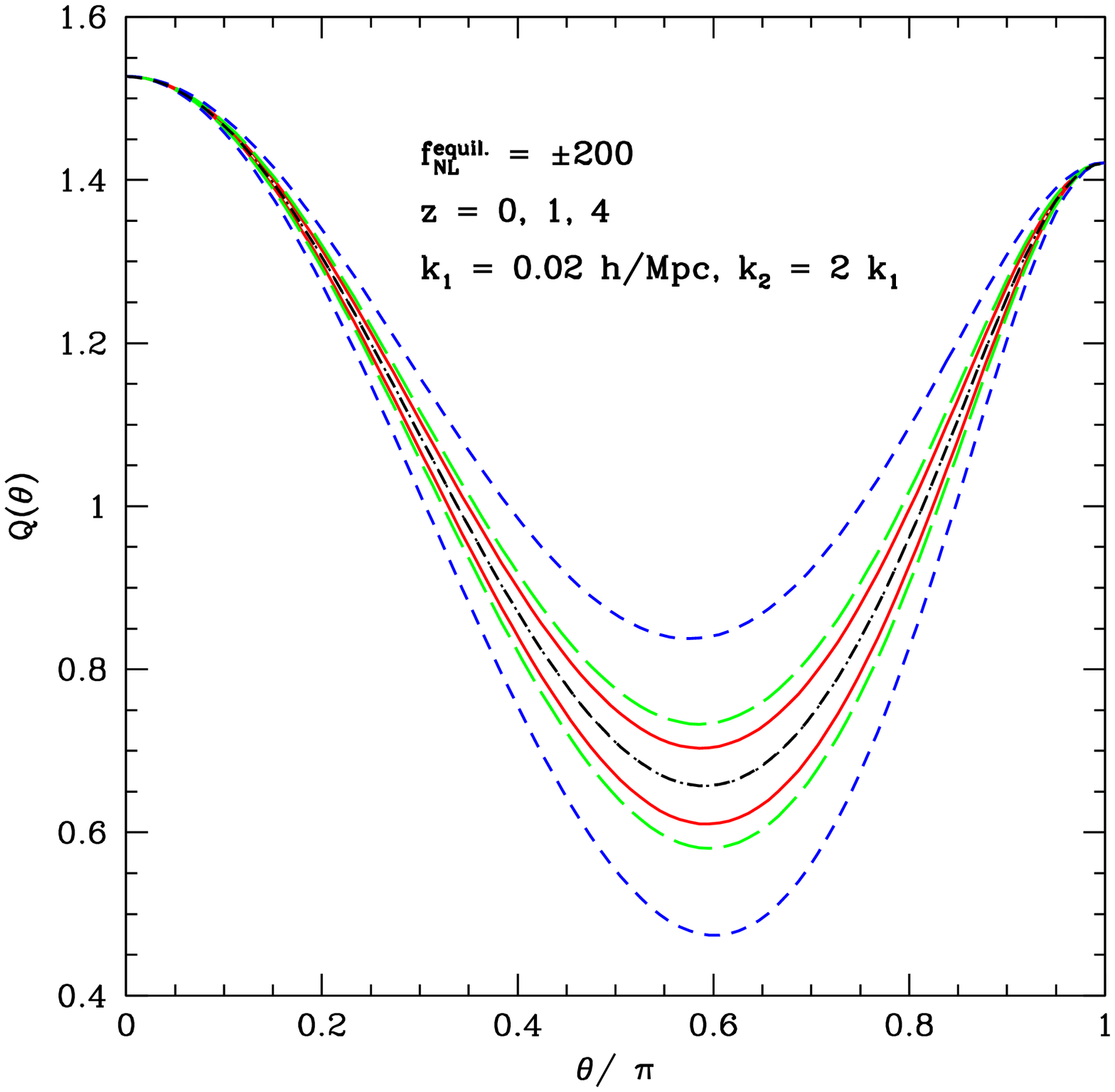} \\
\et
\caption{\label{Fig_NGShapes_an} 
Configuration dependence of the reduced bispectrum of dark matter
distribution from Gaussian and non-Gaussian initial conditions,
as a function of an angle, $\theta$, between two wave vectors,
${\mathbf k}_1$ and ${\mathbf k}_2$, where the magnitude satisfies
 $k_2=2k_1=0.02~\kMpc$ (top panels) and $0.04~\kMpc$ (bottom panels).
({\it Left panels})  $\fNLl=\pm 100$. 
({\it Right panels}) $\fNLe=\pm 200$. 
The dotted black line shows a Gaussian case ($\fNL=0$, redshift
 independent), while the solid, long-dashed and short-dashed  lines show 
non-Gaussian cases at $z=0$, $1$ and $4$, respectively.}
\end{center}
\end{figure*}

How about other configurations? 
In Figure \ref{Fig_NGShapes_an} we plot the  dark matter bispectrum 
with Gaussian or non-Gaussian initial conditions at different redshifts as a 
function of the angle, $\theta$, between ${\mathbf k}_1$ and ${\mathbf
k}_2$, for $k_1=0.01~h~{\rm Mpc}^{-1}$ (top panels) and $0.02~h~{\rm
Mpc}^{-1}$ (bottom panels) and $k_2=2 k_1$.
We find a marked difference in the configuration dependence 
of the two primordial bispectra, local (left panels) and equilateral
(right panels), under consideration.

\subsection{Non-Gaussianity from galaxy bias}

The galaxy bispectrum is most useful for measuring galaxy bias. Assuming that 
galaxy formation is a local process and depends only on the local matter 
density field, one may expand the galaxy number overdensity, $\d_g$, in Taylor 
series of the underlying matter overdensity, $\d$, as \citep{FryGaztanaga1993}
\beq
\d_g(\xv)\simeq b_1\d(\xv)+\frac{1}{2}b_2\d^2(\xv),
\label{localbias}
\eeq
where $b_1$ and $b_2$ are the linear and non-linear bias parameter, 
respectively. It has been shown that this model describes the bispectrum or 
three-point correlation functions from the SDSS and 2dFGRS 
\cite[e.g.,][]{GaztanagaEtal2005,NishimichiEtal2006}, as well as from 
numerical simulations \cite[e.g.,][]{MarinEtal2007}. The galaxy bispectrum is 
given by
\bea
B_g(k_1,k_2,k_3) & \simeq & ~b_1^3 B(k_1,k_2,k_3)
\nonumber \\
& & +b_1^2 b_2[P_L(k_1)P_L(k_2)+{\rm cyc.}],
\eea
up to the second order in matter density fluctuations. Here, $B(k_1,k_2,k_3)$ 
is the intrinsic bispectrum of the underlying matter distribution. The reduced 
galaxy bispectrum is
\beq
Q_g(k_1,k_2,k_3)\simeq \frac{1}{b_1}Q(k_1,k_2,k_3) +\frac{b_2}{b_1^2}.
\label{eq:qg}
\eeq
With $Q(k_1,k_2,k_3)=Q_G(k_1,k_2,k_3)+f_{NL}\widetilde{Q}_I(k_1,k_2,k_3)$, one 
obtains
\bea
Q_g(k_1,k_2,k_3) & \simeq & \frac{1}{b_1}Q_G(k_1,k_2,k_3)
\nonumber\\
& & +\frac{\fNL}{b_1}\widetilde{Q}_I(k_1,k_2,k_3)+\frac{b_2}{b_1^2},
\eea
where we have factorized $f_{NL}$ out from $Q_I$ introducing
$\widetilde{Q}_I\equiv Q_I(\fNL=1)$. 

Finally, the galaxy power spectrum is given by $P_g(k)\simeq b_1^2 P(k)$, up 
to the second order in density fluctuations; however, corrections due to 
non-linear bias appear at the third-order level \citep{HeavensEtal1998,
TaruyaEtal2000,SmithScoccimarroSheth2006,McDonald2006}. Therefore, the bias 
parameters, $b_1$ and $b_2$, affect both the galaxy power spectrum and 
bispectrum. The bispectrum helps us extract the cosmological information from 
the galaxy power spectrum by providing $b_1$ and $b_2$.

\subsection{Redshift space distortion}

The bispectrum measured from redshift surveys is distorted along the line of 
sight direction by radial motion of galaxies. For our analysis in this paper we 
shall deal only with a spherically averaged power spectrum and bispectrum. The 
power spectrum in redshift space after averaging over angles in $k$ space, 
$P_s(k)$, is related to the real space power spectrum by
\beq
P_s(k)=a_0^P(\beta)P_g(k),
\eeq
while the bispectrum is given by
\beq
B_s(k_1,k_2,k_3)=a_0^B(\beta)B_g(k_1,k_2,k_3),
\eeq
where \citep{Kaiser1987,SefusattiEtal2006}
\bea
a_0^P(\beta) & = & 1+\frac{2}{3}\beta+\frac{1}{5}\beta^2, \\ 
a_0^B(\beta) & = & 1+\frac{2}{3}\beta+\frac{1}{9}\beta^2,
\eea
with $\beta\equiv \Omega_m^{5/7}/b_1$.
The reduced bispectrum in redshift space is thus given by
\beq
Q_s(k_1,k_2,k_3)=\frac{a_0^B(\beta)}{[a_0^P(\beta)]^2}
\left[\frac{1}{b_1}Q(k_1,k_2,k_3)+\frac{b_2}{b_1^2}\right].
\label{Qs}
\eeq

We remark that our treatment does not take into account a peculiar 
scale-dependence of redshift distortions that reduces the amplitude of 
non-linear corrections to the reduced matter bispectrum 
\citep{ScoccimarroCouchmanFrieman1999}. This fact only partially justifies the 
approximation of neglecting such corrections altogether.

\section{Fisher matrix analysis}
\label{sec_fisher}
\subsection{Method}

In our analysis we shall consider a set of surveys characterized by their 
volume, $V$, mean galaxy density, $n_g$, and redshift range. We shall assume 
that these surveys have a simple survey geometry, i.e., a contiguous hexahedron.

Our bispectrum estimator is given by \citep{ScoccimarroEtal1998}
\beq
\hat{B} \equiv \frac{V_f}{V_B}
\int_{k_1}\!\!\!\!d^3 q_1\int_{k_2}\!\!\!\!d^3 q_2\int_{k_3}\!\!\!\!d^3 q_3 \;
\delta_D(\qv_{123}) \delta_{\qv_1}\delta_{\qv_2}\delta_{\qv_3},
\label{Best}
\eeq
where the integration is over the bin defined by 
$q_i\in(k_i-\delta k/2,k_i+\delta k/2)$, $V_f=(2\pi)^3/V$ is the volume of the 
fundamental cell in Fourier space, and
\bea
V_B & \equiv & \int_{k_1} \!\!\!\! d^3 q_1\int_{k_2} \!\!\!\! d^3 q_2 
\int_{k_3} \!\!\!\! d^3 q_3 \,\delta_D(\qv_{123})
\nonumber\\
& \simeq & 8\pi^2\ k_1 k_2 k_3\ \Delta k^3,
\eea
with $\Delta k$ a  
multiple of the fundamental frequency, $k_f\equiv 2\pi/L$. We assume that two 
coincide, i.e., $\Delta k=k_f$, thereby taking into account all ``fundamental'' 
triangular configurations.

The variance for our estimator, to the leading order, is given by a triple 
product of the power spectra,
\beq
\Delta B^2 \simeq k_f^3\ \frac{s_{123}}{V_B}\ P(k_1)P(k_2)P(k_3),
\label{Berror}
\eeq
where $s_{123}=6,2,1$ for equilateral, isosceles and general triangles, 
respectively. As for variance of the reduced bispectrum, we assume that 
variance from the bispectrum in the numerator dominates over that from the 
power spectra in the denominator:
\beq
\frac{\Delta Q^2}{Q^2}\simeq\frac{\Delta B^2}{B^2}.
\label{DQ2}
\eeq
We calculate  variance of the redshift-space galaxy reduced bispectrum from 
equations (\ref{DQ2}) and (\ref{Berror}) with $P(k)$ given by
\beq
P_{tot}(k)  \equiv P_s(k)+\frac{1}{(2 \pi)^3} \frac{1}{ \bar{n}},
\eeq
where the second term accounts for the shot noise. We finally obtain
\beq
\Delta Q^2_s(k_1,k_2,k_3)\simeq \frac{s_{123}k_f^3}{V_B}
\frac{P_{tot}(k_1)P_{tot}(k_2)P_{tot}(k_3)}{[P_s(k_1)P_s(k_2)+{\rm cyc.}]^2}.
\label{DQ2B}
\eeq

Once the variance of the reduced bispectrum is given, the Fisher matrix for a 
given redshift bin can be expressed as
\beq
F_{\alpha\beta} \equiv \sum_{k_1,k_2,k_3\le\kMAX}
\frac{\pd  Q_s(i)}{\pd p_{\alpha}}\frac{\pd Q_s(i)}{\pd p_{\beta}}
\frac{1}{\Delta Q_s^2(i)},
\label{fisher}
\eeq
where the parameters, $p_{\alpha}$, represent $b_1$, $b_2$ and $\fNL$. 
We use equation~(\ref{Qs}) for $Q_s$, which is valid only up to the
second order in perturbations. 
Neglecting higher order corrections would introduce
systematic errors when dealing with
the real data, particularly at low $z$ and at small spatial scales,
where non-linearity is substantial and perturbation theory essentially
breaks down. Since we consider high-$z$ surveys on large scales, 
we expect that higher order effects would not  affect our results very much.

As a fiducial cosmological model 
we use a flat $\Lambda$CDM cosmology with matter density $\Omega_m=0.3$, 
baryon density $\Omega_b=0.04$, Hubble parameter $h=0.7$, spectral 
index $n_s=1$ and $\sigma_8=0.9$. 
Of these parameters, $\sigma_8$ and $n_s$ affect our forecast for the
projected errors on the bias parameters most. 
We thus consider also different values such as those
suggested by the  WMAP 3-yr results,
$\sigma_8=0.75$ and $n_s=0.95$ \citep{SpergelEtal2006}.

\subsection{Comments on covariance matrix}
\label{sec_covar}

We shall not include covariance between the cosmological parameters and $b_1$, 
$b_2$, and $f_{NL}$. (We do include covariance between $b_1$, $b_2$, and 
$f_{NL}$.) \citet{SefusattiEtal2006} have shown that, for the SDSS main sample, 
an analysis with the full covariance among all parameters with a prior from the 
WMAP 3-yr results yields an error on $b_1$ that is twice as large as that from 
a simpler analysis without covariance. On the other hand, an error on $b_2$ is 
not affected significantly. Note that the effect on $b_1$ that they observed 
was due mainly to degeneracy between $b_1$ and the amplitude of matter 
fluctuations, as \citet{SefusattiEtal2006} did not use the reduced bispectrum.
We expect that degeneracy would be lifted  in our analysis, as we use the 
reduced bispectrum in which the overall amplitude of matter fluctuations 
cancels. 

More importantly, we shall not include covariance between different triangular 
configurations in the bispectrum. The covariance arises from both 
observational selection functions (i.e., survey geometry and mask) and a 
connected six-point function generated by non-linear gravitational evolution. 
This is a rather crude approximation. 
\citet{ScoccimarroSefusattiZaldarriaga2004} have included the full reduced 
bispectrum covariance plus the peculiar survey geometry, when they calculate 
the constraints on galaxy bias and primordial non-Gaussianity from the SDSS 
main sample. Using the same realizations of the survey and the same estimator 
for the covariance matrix, \citet{SefusattiScoccimarro2005} have compared the 
analysis with the full covariance matrix (including the observational 
selection function) and that with an approximate diagonal Gaussian variance. 
They have found that the latter simplified treatment overestimates the 
signal-to-noise by a factor of 2 for $\kMAX\sim 0.1 \kMpc$, and a factor of 8 
for $\kMAX\sim 0.3 \kMpc$ at redshift zero. It is not clear, however, how to 
separate the contribution from non-linear evolution from the effect of the 
selection function. One would generically expect the radial contribution to be 
smaller at high $z$, as the six-point function from gravitational clustering 
becomes smaller than the non-connected part of six-point function (which 
consists of power spectra) at higher $z$. In any case, our results should be 
taken as a guide, and one needs to perform the full analysis including the 
selection functions peculiar to a given survey design.

\subsection{Non-linearity and maximum wavenumber}

While one can measure the galaxy power spectrum or bispectrum down to very 
small spatial scales, say, 10~kpc, it is challenging to extract useful 
cosmological information from such small spatial scales owing to strong 
non-linearity. Therefore, one has to decide on the maximum wavenumber, $\kMAX$,
below which theory may be trusted. Not surprisingly, since there are many more 
modes available on smaller spatial scales, the amount of cosmological 
information one can extract from data grows as $\kMAX$ increases. It is 
therefore important to use a realistic $\kMAX$ in order not to overestimate the
statistical power of a given galaxy survey design.

How do we decide on $\kMAX$? The first obvious thing to do would be to test our
theory of the power spectrum and bispectrum against numerical simulations. A 
value of $\kMAX$ can be found by comparing perturbation theory predictions with
numerical simulations \citep[see e.g.,][for an analysis for the matter power
spectrum]{JeongKomatsu2006}. It is likely that a simple model provided by the 
second-order (tree-level) bispectrum given by equation (\ref{Qs}) breaks down 
at a relatively small $k$ due to non-linearities of gravitational growth as 
well as due to non-linear or even non-local bias. Therefore, a model of the 
bispectrum that takes into account higher-order perturbations would be 
necessary to push $\kMAX$ further. New promising techniques such as a 
renormalized perturbation theory approach \citep{CrocceScoccimarro2006a,
CrocceScoccimarro2006b,CrocceScoccimarro2007,Mcdonald2007,
MatarresePietroni2007} may be used to obtain better predictions for the power 
spectrum and bispectrum. Further progress is required particularly for 
understanding redshift distortions \citep{Scoccimarro2004}.

In this paper we use a very simple prescription for getting $\kMAX$. We choose 
$\kMAX$ so that $\sigma(R_{\rm min},z)= 0.5$ and $\kMAX=\pi/(2R_{\rm min})$. 
The main motivation for this choice being that for small perturbations in the 
matter distribution, say, $\sigma(R,z)<1$, one may reasonably expect that an 
analytical model for non-linearities is viable. Note that $\kMAX$ derived in 
this way depends on $z$, as $\sigma(R,z)=\sigma(R,0)D(z)/D(0)$.

An alternative, much more conservative estimate of $\kMAX$ could be given by 
requiring that the error on $b_1$ derived from the tree-level bispectrum for 
some $\kMAX$ does not exceed the higher-order (``1-loop'', or 4th-order 
perturbation) corrections in perturbation theory to the reduced matter 
bispectrum at the same $\kMAX$. Specifically, one may use
\beq
\frac{\Delta b_1}{b_1}\ge
\frac{\Delta Q_{eq}^{\rm 1-loop}(\kMAX)}{Q_{eq}^{\rm tree}(\kMAX)},
\label{kMAXb}
\eeq
to determine $\kMAX$. Here, $\Delta b_1$ is computed including all scales down 
to $\kMAX$ and $\Delta Q^{\rm 1-loop}(\kMAX)$ is the 1-loop corrections 
\citep{Scoccimarro1997,ScoccimarroEtal1998} to the tree-level reduced 
bispectrum, $Q^{\rm tree}$, evaluated for equilateral configurations with 
$k_1=k_2=k_3=\kMAX$. This approach, however, makes no use of a large amount of 
information on small scales, and is far from being optimal. Also, the 1-loop 
correction to the bispectrum, as it is the case for the power spectrum, tends 
to overestimate the non-linear behaviour measured in simulations, thereby 
making this approach even more conservative than necessary. In 
Sec.~\ref{sec_volume} we shall compare these two approaches  as a function of 
volume and number density. For the expressions of the 1-loop corrections to the 
reduced bispectrum see, e.g., \citet{BernardeauEtal2002}.

\subsection{Fiducial values for the galaxy bias parameters}
\label{sec_bias}

The galaxy bias parameters, $b_1$ and $b_2$, depend on a number of factors, 
including galaxy populations, luminosities, and redshifts. On the other hand, 
the bias of dark matter halos, which can be calculated from $N$-body 
simulations, is understood relatively well. Therefore, the galaxy bias can be 
calculated from the dark matter halo bias, if we assume that galaxies form in 
dark matter halos. To do this, one needs (at least) the following information: 
(i) the halo bias \citep{MoJingWhite1997,ShethTormen1999}, and (ii) how each 
halo is populated with galaxies, that is, the  Halo Occupation Distribution 
(HOD), $\langle N\rangle_M$.
 
\begin{figure*}[t]
\begin{center}
\includegraphics[width=0.45\textwidth]{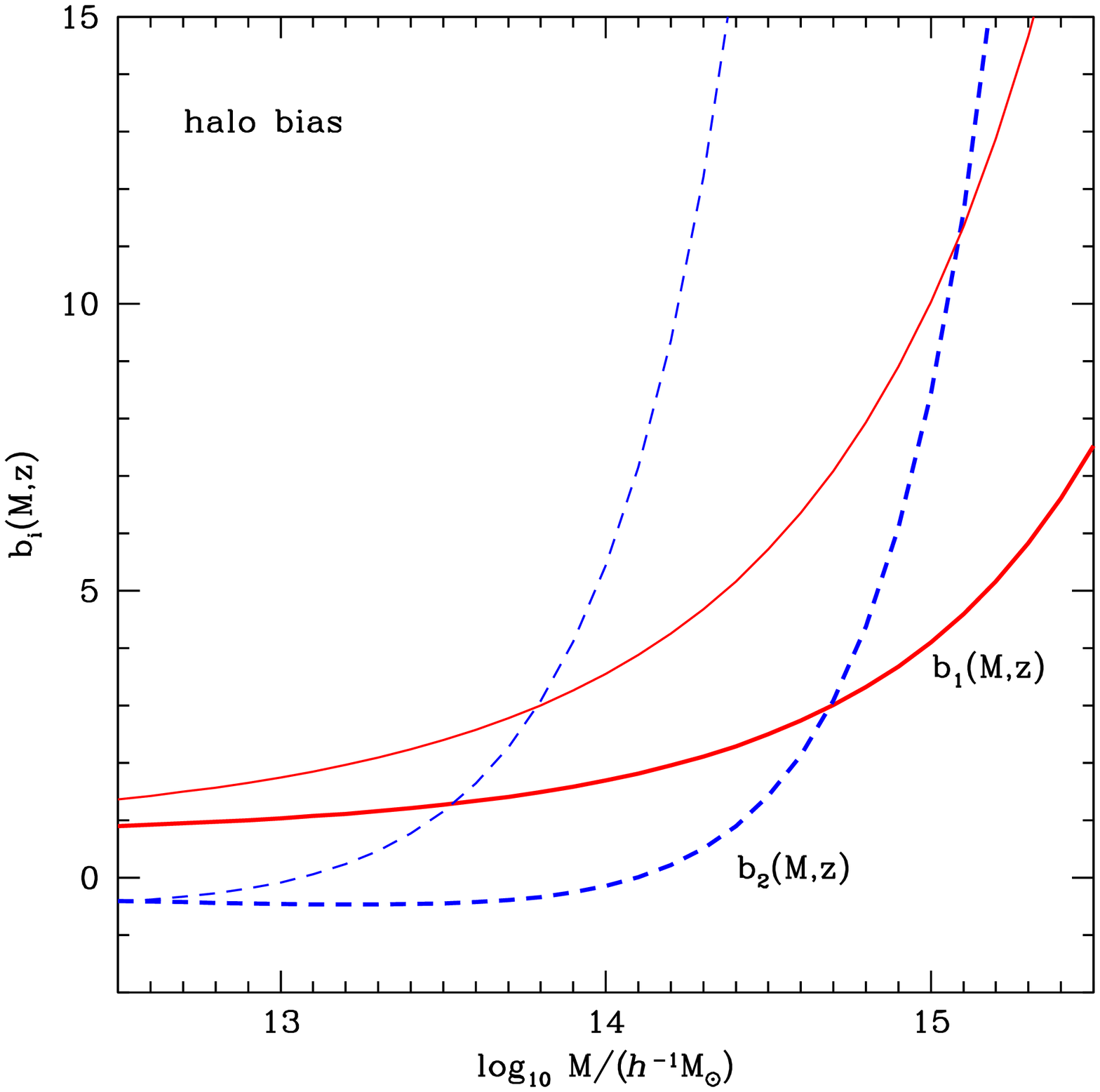}
\includegraphics[width=0.45\textwidth]{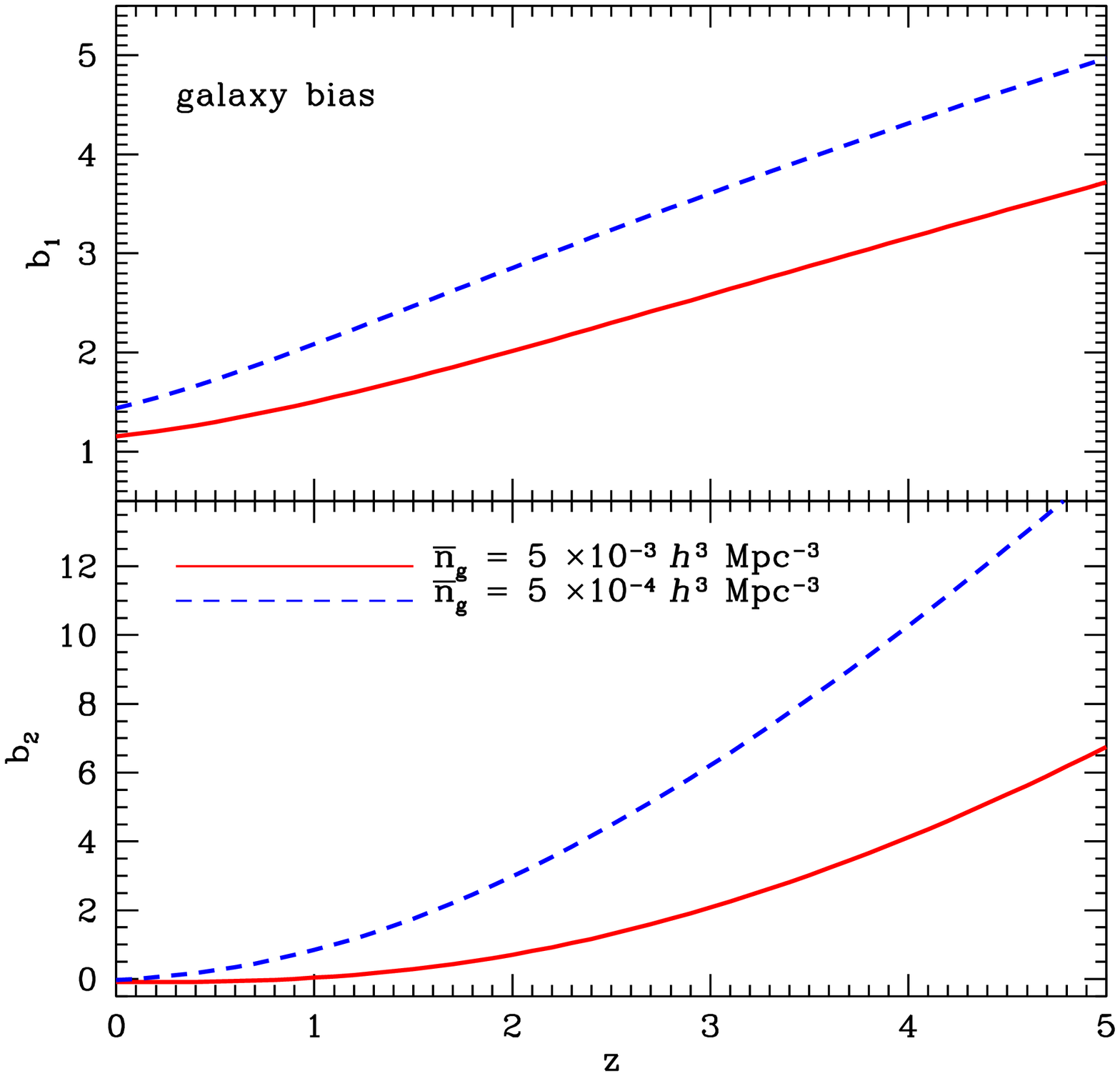}
\caption{\label{fig_bias} 
({\it Left panel}) The halo bias functions, $b^h_1(M,z)$ (solid lines) and 
$b^h_2(M,z)$ (dashed lines), as a function of the mass, $M$, for $z=0$ (thick 
lines) and $z=1$ (thin lines) in the approximation that the formation redshift 
equals the observation redshift, $z=z_f$. ({\it Right panel}) The galaxy bias 
parameters, $b_1$ and $b_2$, for the mean galaxy density of 
$n_g=5\times 10^{-3}$ (continuous lines) and 
$n_g=5\times 10^{-4}~h^3~{\rm Mpc}^{-3}$ (dashed lines)}.
\end{center}
\end{figure*}

We calculate the galaxy bias parameters from the large scales expression
\beq\label{biasHOD}
b_i\simeq\frac{1}{n_g}\int_{M_{min}}\!\!\!dM n_h(M,z)~ b^h_i(M,z) \langle N
\rangle_M,
\eeq
for $i=1$ and $2$, where $n_h(M,z)$ is the mass function of dark matter halos 
of mass $M$ at redshift $z$, $b^h_i(M,z)$ is the halo bias function, and the 
HOD, $\langle N\rangle_M$, is the mean number of galaxies per halo of a given 
mass, $M$. We shall use the Sheth \& Tormen's formula for $n_h(M,z)$  
\citep{ShethTormen1999}:
\beq
n_h(M,z)=-\frac{\bar{\rho}}{M^2}\frac{d\ln\sigma}{d\ln M}f(\nu)
\eeq
where $\nu=\delta_c/\sigma(M,z)$ with $\delta_c=1.686$, and
\beq
f(\nu)=A\sqrt{\frac{2q}{\pi}}\left[1+(q\nu^2)^{-p}\right]\nu e^{-q\nu^2/2}
\eeq
with $A=0.322$, $p=0.3$ and $q=0.707$. The halo bias parameters, $b_1^h$ and 
$b_2^h$, are given by \citep{MoJingWhite1997,ScoccimarroEtal2001A}
\bea\label{halo_bias}
b^h_1(M,z|z_f) & = & 1 + \ep_1+E_1, \nonumber\\
b^h_2(M,z|z_f) & = & \frac{8}{21}(\ep_1+E_1)+\ep_2+E_2,
\eea
where $z$ refers to the redshift of observation, while $z_f$ refers to the 
redshift of formation of halos of mass $M$, and
\beq
\ep_1 = \frac{q\nu^2-1}{\d_f},\qquad
\ep_2 = \frac{q\nu^2}{\d_f}\frac{q\nu^2-3}{\d_f},
\eeq
\beq
E_1 = \frac{2p/\d_f}{1+(q\nu^2)^p},\qquad
\frac{E_2}{E_1} = \frac{1+2p}{\d_f}+2\ep_1,
\eeq
with $\d_f=\d_cD(z)/D(z_f)$, where $D(z)$ is the linear growth function. In the 
left panel of Figure \ref{fig_bias} we show the halo bias functions, $b_1(M,z)$
and $b_2(M,z)$, as a function of $M$ and $z$, in the approximation that the 
formation redshift equals the observation redshift, $z=z_f$. We shall always 
assume this throughout the paper.

As for the HOD, we adopt the form proposed by  \citet{TinkerEtal2005}:
\beq
\langle N\rangle_M=1+\frac{M}{M_1}\exp\left(-\frac{M_{cut}}{M}\right)
\eeq
for $M>M_{min}$ and zero otherwise. The parameter $M_{min}$ represents the 
minimum mass above which we find a (central) galaxy in the halo, while $M_1$ 
represents the mass above which we can find a second (satellite) galaxy. 
Measuring the HOD parameters for subhalo populations from several N-body 
simulations at different redshifts and densities, 
\citet{ConroyWechslerKravtsov2006} found a correlation between $M_{cut}$ and 
$M_1$ given by
\beq\label{Mcut}
\log_{10} M_{cut}=0.76\log_{10} M_1+2.3.
\eeq
One also finds from Table 2 in \citet{ConroyWechslerKravtsov2006} that 
$M_1/M_{min}$ depends on redshift and density only weakly; thus, for simplicity 
we shall keep this ratio fixed at $\log_{10}(M_1/M_{min})=1.1$, and find 
$M_{min}$ from 
\beq
n_g=\int_{M_{min}}dM n_h(M,z) \langle N\rangle_M,
\eeq
for a given $n_g$. 

In the right panel of Figure \ref{fig_bias} we show the galaxy bias 
parameters, $b_1$ and $b_2$, from equation (\ref{biasHOD}) as a function of 
redshift for two values of the mean galaxy density, $n_g=5\times 10^{-3}$ and 
$5\times 10^{-4}~h^3~{\rm Mpc}^{-3}$. As expected, for a fixed galaxy number 
density the value of the linear bias, $b_1$, increases with redshift. We find 
that $b_1$ and $b_2$ are strongly correlated. We shall come back to this point 
in Sec.~\ref{sec_hod}.

We admit that these values are derived from very simplified models without 
much justification. We need observational data to determine the true bias 
parameters for high-$z$ surveys eventually, although we do not have sufficient 
data for doing so yet. Nevertheless, we find our approach useful for our 
purpose of deriving the fiducial values of $b_1$ and $b_2$ with a realistic 
redshift evolution, particularly for redshift surveys spanning a wide range in 
redshift, for which one has to consider a set of redshift bins and assume 
different fiducial values for $b_1$ and $b_2$ at different $z$. We note that 
$b_1$ obtained from our method agrees with those obtained in the previous work 
by assuming $\sigma_{8,g}\simeq 1$ and $n_g\simeq  5\times 10^{-4}\icMpc$
 \citep{SeoEisenstein2003,TakadaKomatsuFutamase2006}.  

In Sec.~\ref{sec_hod} we shall show how one can extend this simple picture by 
introducing a redshift dependence in the HOD, and how one can make use of the 
information on galaxy bias derived from the bispectrum to constrain the HOD 
parameters directly.

\section{Results}
\label{sec_results}

In this section we present the results from our Fisher matrix analysis of the 
galaxy bispectrum. We first study how the derived constraints on the galaxy 
bias parameters and primordial non-Gaussianity depend on the choice of 
$\kMAX$, taking into account the two approaches discussed above. We then study 
how the constraints depend on the survey volume and redshift. Finally we shall 
apply our method to make forecasts for several current and proposed redshift 
surveys.

\subsection{Dependence on $\kMAX$, volume, redshift  and number density}

\subsubsection{$\kMAX$}

As mentioned in the previous section, constraints on galaxy bias and 
primordial non-Gaussianity would depend strongly on  $\kMAX$, the smallest 
scale included in the analysis. 

As an example, we consider sample surveys at  two redshifts: the median 
redshifts of (i) $\bar{z}=1$ and (ii) $\bar{z}=3$. Each has the volume of 
$V=10\cGpc$ and the number density of $n_g=5\times 10^{-3}\icMpc$. (The total 
number of galaxies in the survey volume at each redshift is 50 million 
galaxies.) The bias parameters are (i) $b_1=1.5$ and $b_2=0.035$ at 
$\bar{z}=1$, and (ii) $b_1=2.6$ and $b_2=2.1$ at $\bar{z}=3$.

\begin{figure*}[t]
\begin{center}
\includegraphics[width=0.85\textwidth]{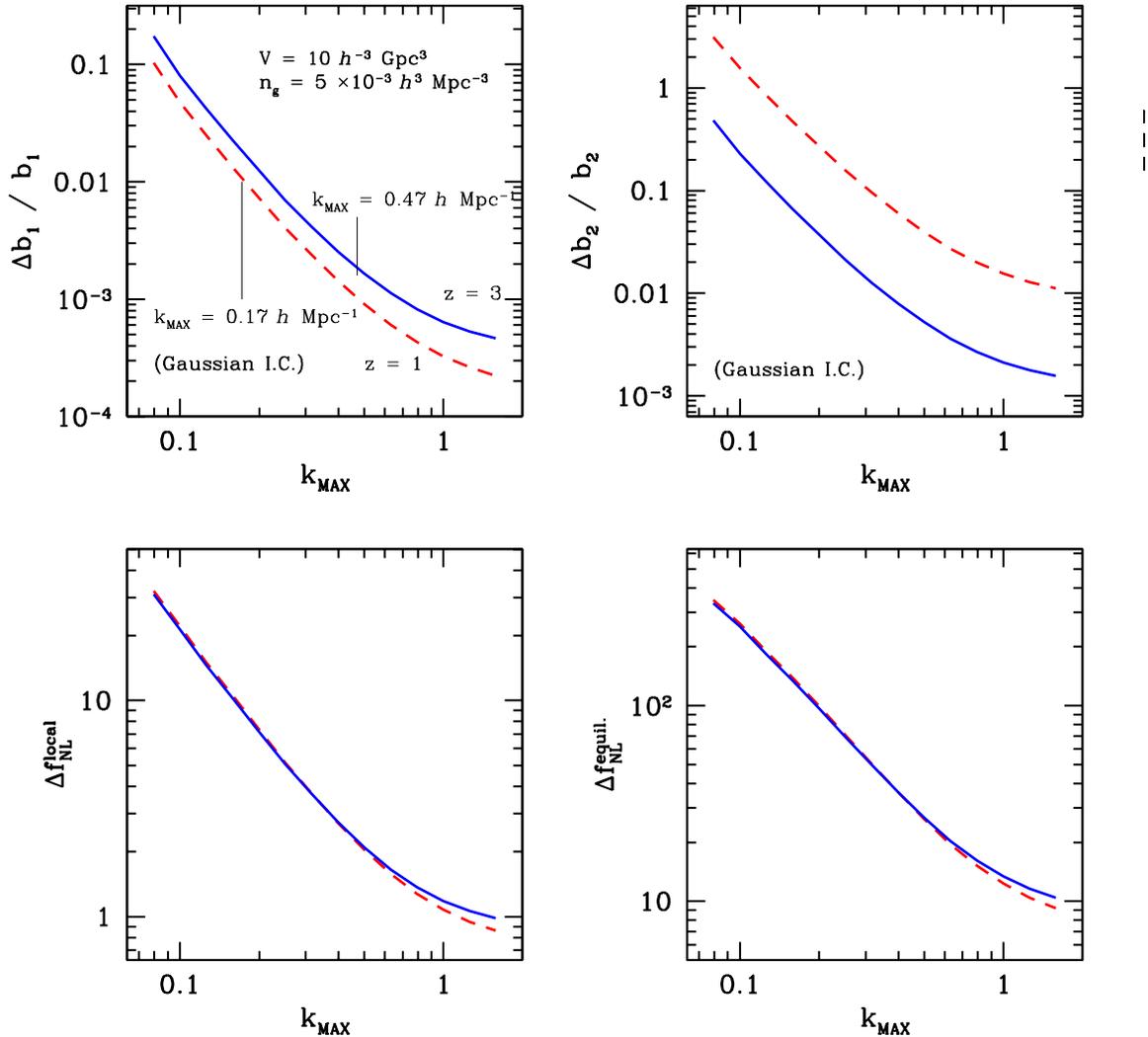}
\caption{\label{fig_scalings_kmax_dm3} ({\it Upper panels}) 
Predicted errors on galaxy bias parameters vs the maximum wavenumber, $\kMAX$.
The dashed and solid lines show the prediction for a galaxy survey  at $z=1$ 
and 3, respectively. Each survey is assumed to have the survey volume of 
$V=10\cGpc$ and the number density of $n_g=5\times 10^{-3}\icMpc$. The left 
panel shows the marginalized 1-$\sigma$ errors on the linear bias, $b_1$, 
while the right panel shows the non-linear bias, $b_2$. Both assume Gaussian 
initial conditions, $\fNL=0$. The vertical lines show  $\kMAX$ as determined 
from  $\sigma(R;z) = 0.5$ for each redshift (see Sec.~3.2). ({\it Lower 
panels}) Predicted errors on primordial non-Gaussian parameters vs $\kMAX$.
The left panel shows the marginalized 1-$\sigma$ errors on the local model,
 $\fNLl$, while the right panel shows the equilateral model, $\fNLe$. The bias 
parameters have been marginalized.}
\end{center}
\end{figure*}

In the upper panels of Figure~\ref{fig_scalings_kmax_dm3} we plot the 
marginalized, 1-$\sigma$, fractional errors on $b_1$ and $b_2$ at  $\bar{z}=1$ 
and 3, assuming Gaussian initial conditions, i.e., $\fNL=0$. We observe an 
interesting effect: a fractional error on $b_1$ improves at lower $z$, while 
that on $b_2$ improves at higher $z$. (Note that this statement is true only 
when {\it the same $\kMAX$ is used at both redshifts}. See discussion below.)
This can be understood as follows. Let us recall the form of the galaxy
reduced bispectrum (Eq.~[\ref{eq:qg}]):
$$
Q_g(k_1,k_2,k_3)\simeq \frac{1}{b_1}Q(k_1,k_2,k_3) +\frac{b_2}{b_1^2}.
$$
Now, $Q$ on the right hand side is independent of $z$ at the tree-level when 
initial fluctuations are Gaussian. Therefore, the first term falls as $1/b_1$ 
at higher $z$ where $b_1$ is larger (Fig.~\ref{fig_bias}). On the other hand, 
the second term actually grows as $z$: for the current example 
$b_2/b_1^2=0.016$ at $z=1$ and 0.31 at $z=3$. Therefore, our sensitivity to 
$b_2$ grows with $z$, while our sensitivity to $b_1$ declines with $z$.

Let us study more quantitatively the sensitivity to $b_1$. In the limit of 
linear bias, $b_2=0$, a signal-to-noise of the reduced bispectrum of 
equilateral configurations is given by
\bea
\left.\frac{Q_s^2(k)}{\Delta Q_s^2(k)}\right|_{b_2=0} & \simeq &
\frac{[a_0^B(\beta)]^2}{[a_0^P(\beta)]^3}
\frac{V_B}{k_f s_{123}}\frac{B_G^2(k,k,k;z)}{P^3_L(k;z)}
\nonumber\\
& \propto & D^2(z).
\label{stn_b1}
\eea
We expect, therefore, that a signal-to-noise of the bispectrum from
 gravitational instability declines with $z$, resulting in an increasing error 
on $b_1$ at higher $z$.

In practice, however, we predict that galaxy surveys at higher $z$ should
result in better determinations of both $b_1$ and $b_2$. The reason is quite 
simple: $\kMAX$ at higher $z$ must be larger than that at lower $z$. In the 
upper left panel of Figure \ref{fig_scalings_kmax_dm3} we show $\kMAX$ as 
determined from $\sigma(R;z)= 0.5$: $\kMAX=0.17\kMpc$ at $z=1$ and 
$\kMAX=0.47\kMpc$ at $z=3$. The difference is clear: when the modes up to 
$\kMAX$ are included, a survey at $z=3$ yields an error on $b_1$ that is a 
factor of 5 better than that at $z=1$. As for $b_2$, a survey at $z=3$ does 
better by nearly two orders of magnitude.

\begin{figure*}[t]
\begin{center}
\includegraphics[width=0.85\textwidth]{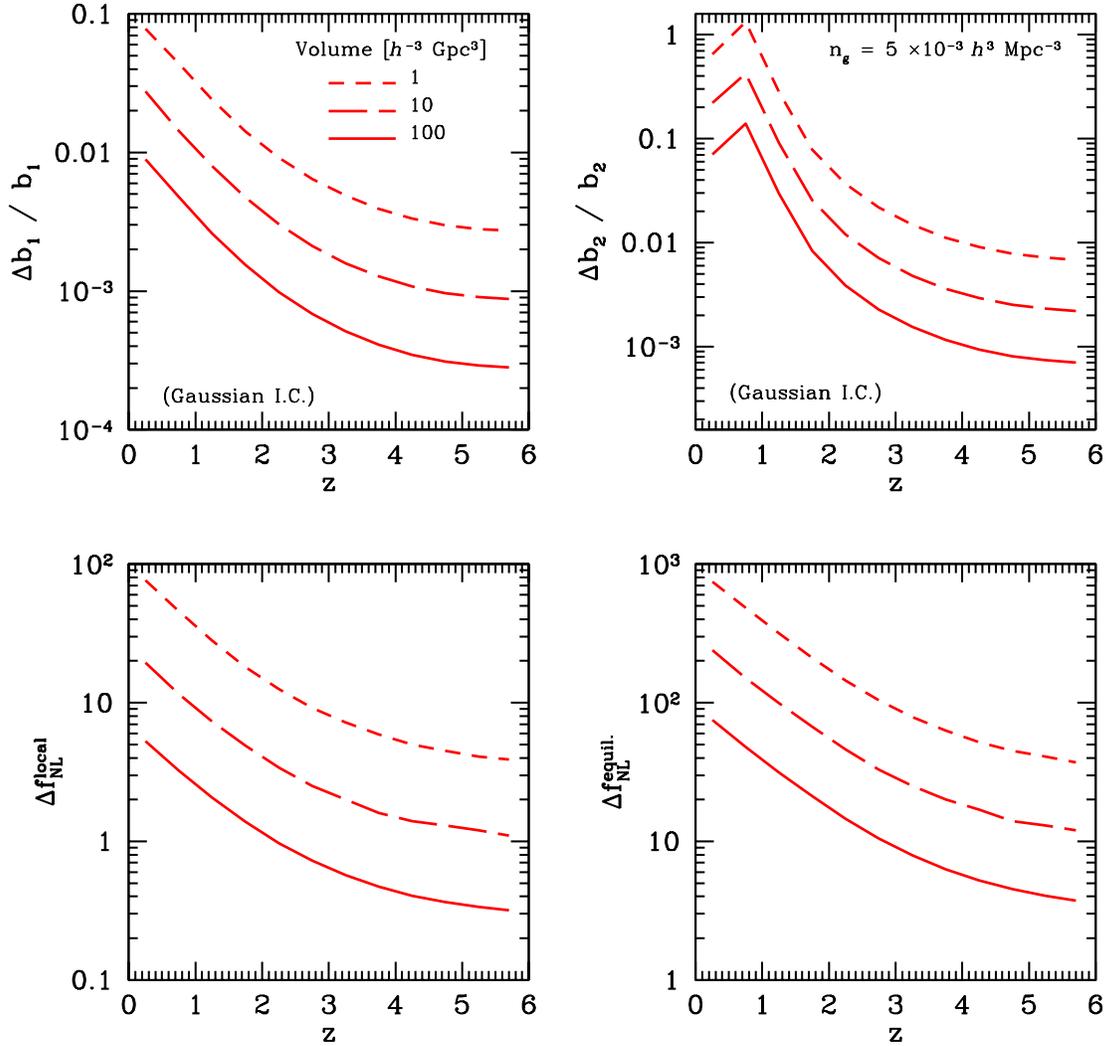}
\caption{\label{fig_scalings_dm3} 
Predicted  1-$\sigma$ errors on galaxy bias and primordial non-Gaussianity vs 
the survey volume, $V$, and redshift, $z$. The short-dashed, long-dashed and 
solid lines show $V=1$, $10$ and $100\cGpc$, respectively, with the galaxy 
number density of $n_g=5\times 10^{-3}\icMpc$. ({\it Upper panels}) Fractional 
errors on the linear bias, $b_1$ (left), and non-linear bias, $b_2$ (right), 
for Gaussian initial conditions, $\fNL=0$. ({\it Lower panels}) Errors on 
primordial non-Gaussian parameters,  $\fNLl$ (left) and $\fNLe$ (right), 
marginalized over $b_1$ and $b_2$.} 
\end{center}
\end{figure*}

How about primordial non-Gaussianity? In the lower panels of Figure 
\ref{fig_scalings_kmax_dm3} we show the predicted errors on $\fNLl$ and 
$\fNLe$, marginalized over $b_1$ and $b_2$. We find that the difference 
between $z=1$ and 3 is negligible at the same $\kMAX$. This is a consequence 
of the fact that a signal-to-noise for the primordial bispectrum component is 
not, in the first approximation, redshift dependent. For equilateral 
configurations one finds 
\bea
\left.\frac{Q_s^2(k)}{\Delta Q_s^2(k)}\right|_{I} & \simeq &
\frac{[a_0^B(\beta)]^2}{[a_0^P(\beta)]^3}
\frac{V_B}{k_f s_{123}}\frac{B_I^2(k,k,k;z)}{P^3_L(k;z)}
\nonumber\\
& \propto & {\rm constant}
\label{stn_initial}
\eea
where we considered only the primordial term, $Q_I$. Nevertheless, we still 
predict that galaxy surveys at higher $z$ should result in better 
determinations of both $\fNLl$ and $\fNLe$, as $\kMAX$ must be larger at 
higher $z$ and therefore many more modes are available for the analysis at 
higher $z$.

Results from a much more conservative estimate of $\kMAX$ (eq.~[\ref{kMAXb}]) 
will be given near the end of the next section.

\subsubsection{Volume, redshift, and number density of galaxies}
\label{sec_volume}

Dependence of the predicted errors on volume is straightforward: it depends 
simply on $1/\sqrt{V}$. Dependence on $z$ is a combination of two effects: (i) 
how a signal-to-noise for a given $\kMAX$ grows with $z$, and (ii) how $\kMAX$ 
grows with $z$. Finally,  the number density of galaxies determines a 
signal-to-noise on small scales, where the shot noise plays an important role. 
In particular, very high-$z$ surveys at, e.g., $z\gtrsim 3$, do not add very 
much if the number density of galaxies is too low.

In Figure \ref{fig_scalings_dm3} we show how the predicted constraints on the
galaxy bias parameters, $b_1$ and $b_2$, improve with the survey volume, as a 
function of the median redshift, $\bar{z}$. We used 
$n_g=5\times 10^{-3}\icMpc$ for the number density of galaxies. The fiducial 
values of $b_1$ and $b_2$ are calculated for each $\bar{z}$ from Figure 
\ref{fig_bias}. We used  $\kMAX$ determined from $\sigma(R,\bar{z})= 0.5$ for 
a given $\bar{z}$. Since $\kMAX$ grows as $\bar{z}$ increases, the predicted
constraints on $b_1$ and $b_2$ also improve as $\bar{z}$ increases. A spike at 
$z\sim 0.8$ in $\Delta b_2/b_2$ is a numerical artifact of $b_2$ being very 
close to zero. The dependence on volume is given simply by $1/\sqrt{V}$.

\begin{figure*}[t]
\begin{center}
\includegraphics[width=0.85\textwidth]{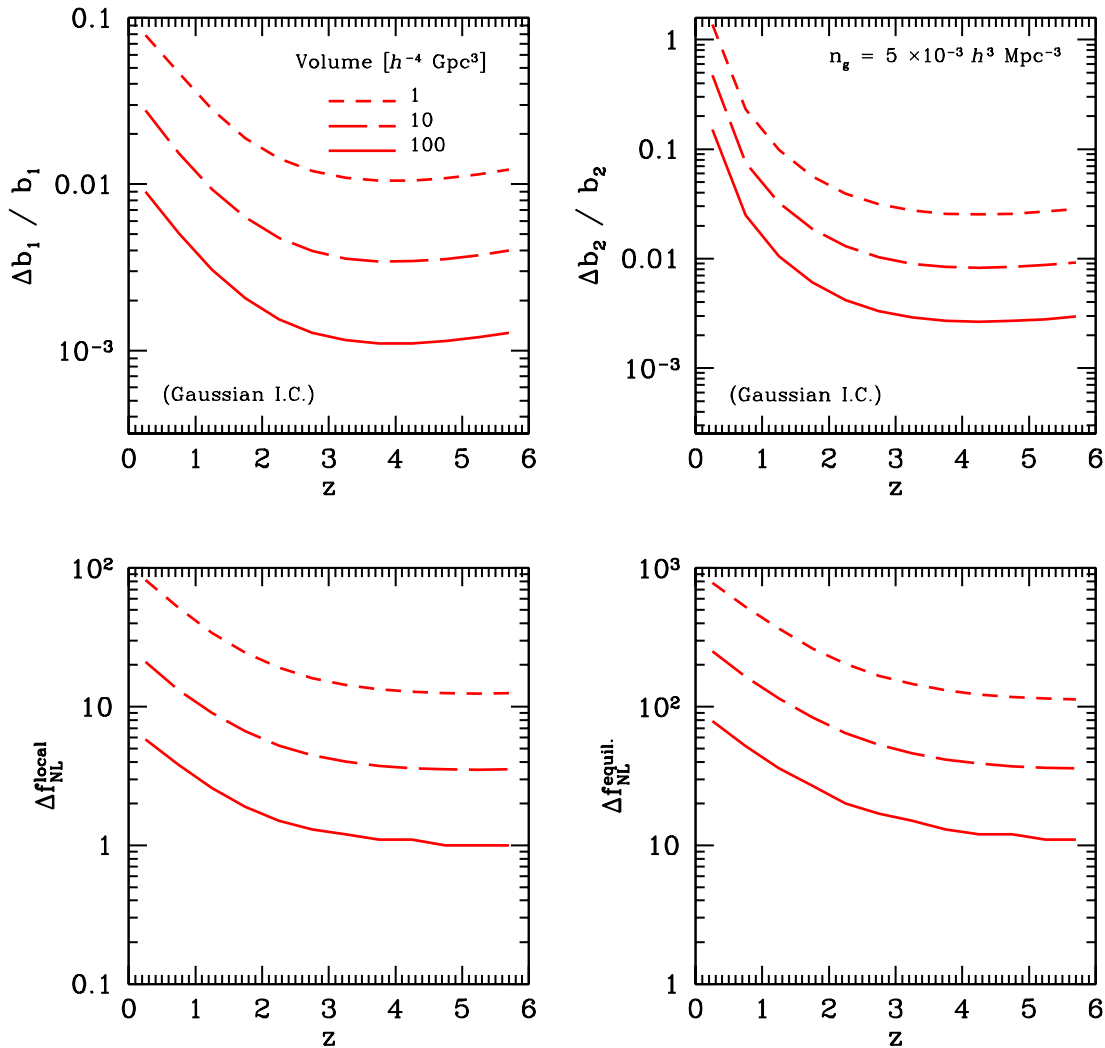}
\caption{\label{fig_scalings_dm4} Same as Figure \ref{fig_scalings_dm3}
 but for a smaller number density of galaxies, $n_g=5\times
 10^{-4}\icMpc$.}
\end{center}
\end{figure*}

In the lower panels of Figure \ref{fig_scalings_dm3} we show the predicted 
constraints on primordial non-Gaussianity, marginalized over $b_1$ and $b_2$. 
We find that a survey of the size $V\sim 1\cGpc$ at $z\sim 4-6$ or 
$V\sim 10\cGpc$ at $z\sim 1-2$ is as sensitive to $\fNLl$ as the CMB data from 
Planck. A more ambitious design, e.g., $V\sim 10\cGpc$ at $z\gtrsim 2$, can 
achieve $\Delta \fNLl\sim 1$, although it depends on the number density quite 
strongly. An even more ambitious design, $V\sim 100\cGpc$, would enable us to 
detect the primordial bispectrum from ubiquitous non-Gaussianity ``floor'' 
from the second-order evolution of primordial fluctuations 
\citep{BartoloMatarreseRiotto2005}.

Constraints on the equilateral type of non-Gaussianity suffer from a stronger 
degeneracy between the primordial bispectrum and the non-linear gravitational 
evolution as well as non-linear bias, and thus the predicted errors on $\fNLe$ 
are an order of magnitude larger than those on $\fNLl$. (See Sec.~\ref{sec:deg}
for more detail.) Nevertheless, a survey of $V\sim 10\cGpc$ at $z\sim 1$ 
should provide a constraint that is comparable to that from the WMAP 3-yr data.
 
We also computed the Fisher matrix for an all-sky survey from $z=0$ to $5$. We 
divided the entire redshift range in bins of the size $\Delta z=0.5$, and used 
$n_g=5\times 10^{-3}\icMpc$. We find that such a survey should provide
$\dfNLl\sim 0.2$  and $\dfNLe\sim 2$. These values probably represent the best
limits on $\fNL$ one can ever hope to achieve from galaxy surveys.

How about the number density of galaxies? When the number density is low, the 
shot noise completely dominates at small scales, and thus one fails to improve 
a signal-to-noise by increasing  $\kMAX$. This suggests that very high-$z$ 
galaxy surveys do not add much if the number density of galaxies is too low. 
In Figure \ref{fig_scalings_dm4} we show the case for $n_g=5\times 10^{-4}
\icMpc$. Clearly, our sensitivity to all of $b_1$, $b_2$, $\fNLl$ and $\fNLe$ 
does not improve at all beyond $z\sim 3$. Therefore, it  makes sense to 
conduct very high-$z$ surveys, only if one can detect more than $n_g\sim 
10^{-3}\icMpc$.

\begin{figure*}[t]
\begin{center}
\includegraphics[width=0.85\textwidth]{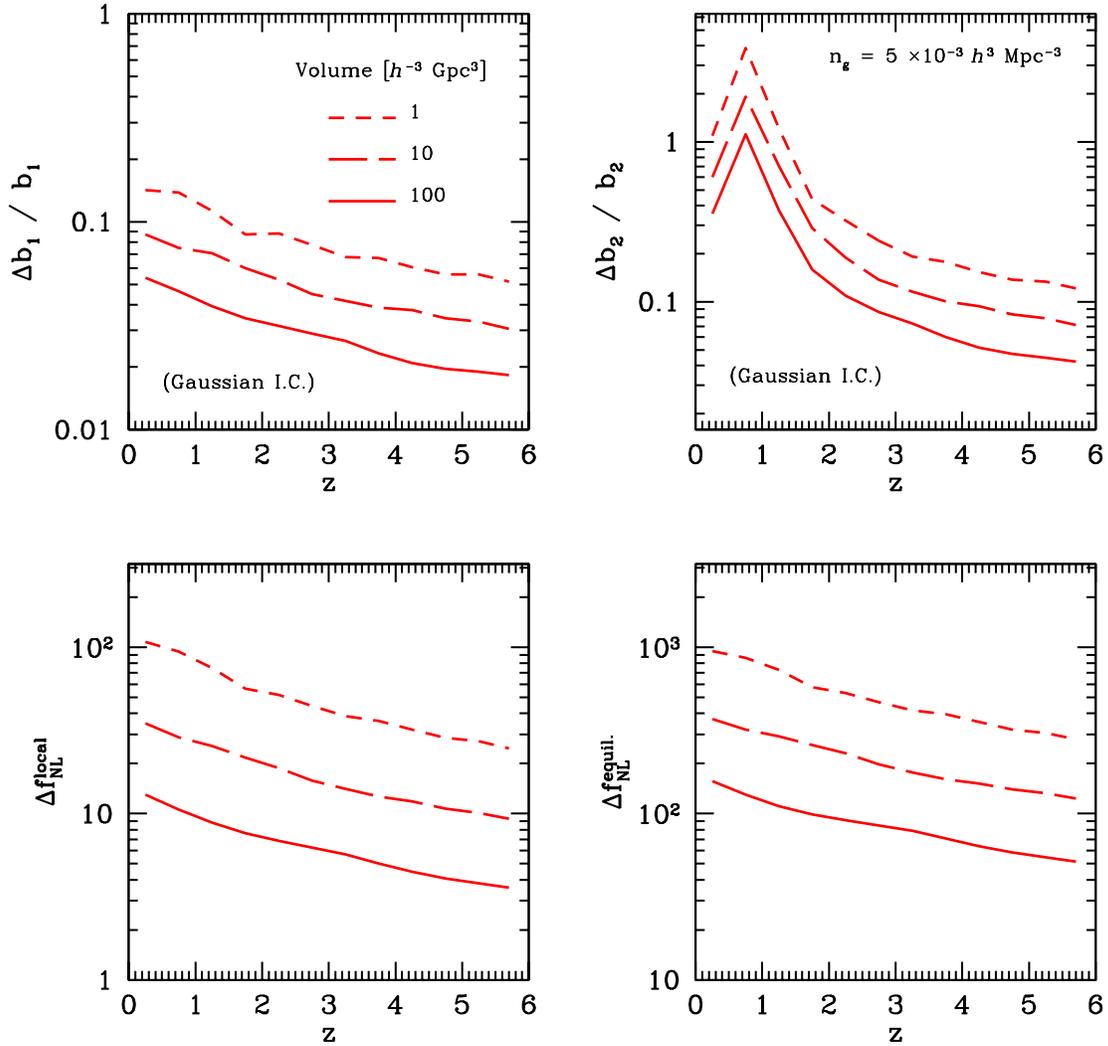}
\caption{\label{fig_scalings_dm3_kB} Same as fig.
~\ref{fig_scalings_dm3} but for a much more conservative estimate of 
$\kMAX$ given by equation (\ref{kMAXb}).}
\end{center}
\end{figure*}

How robust are these results? The most uncertain parameter in our analysis is 
$\kMAX$. What if $\kMAX$ is significantly lower than that from $\sigma(R,z)=
0.5$? To address this question, we have repeated our analysis using a much 
more conservative estimate of $\kMAX$ given by equation (\ref{kMAXb}). 
This estimate was derived by throwing away any information beyond $\kMAX$ at 
which the tree-level bispectrum becomes inaccurate. This is a conservative 
estimate because we can certainly improve our theoretical prediction by going 
to the higher order, ``1-loop'' (4th order) calculations 
\citep{Scoccimarro1997,ScoccimarroEtal1998}. We show the results in Figure 
\ref{fig_scalings_dm3_kB}. We find  significantly weaker constraints; for 
example, fractional errors on $b_1$ from a survey of  $V=10\cGpc$ now go from 
$9\%$ to a few percent  from $z=0$ to $z=5$, or a factor of $\sim 8$ and 20 
weaker constraints at  $z\sim 1$ and 5, respectively. For this choice of 
$\kMAX$ the shot-noise is not very important because we are restricted to a 
fairly small $k$ already. Therefore we obtain similar results for a lower 
density, $n_g=5\times 10^{-4}\icMpc$.

\subsection{Parameter degeneracy}
\label{sec:deg}

\begin{figure*}[t]
\begin{center}
\includegraphics[width=0.45\textwidth]{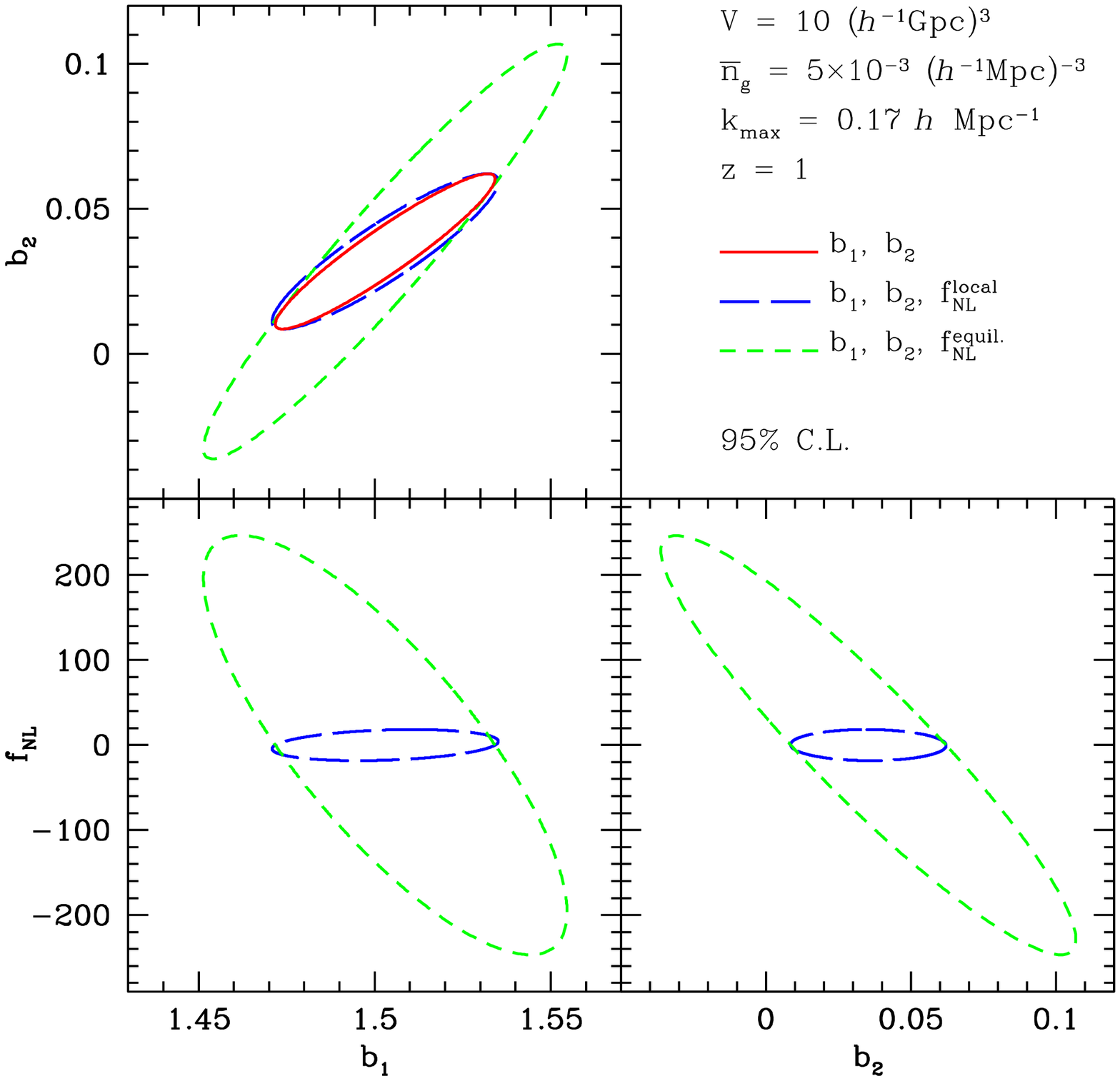}
\includegraphics[width=0.45\textwidth]{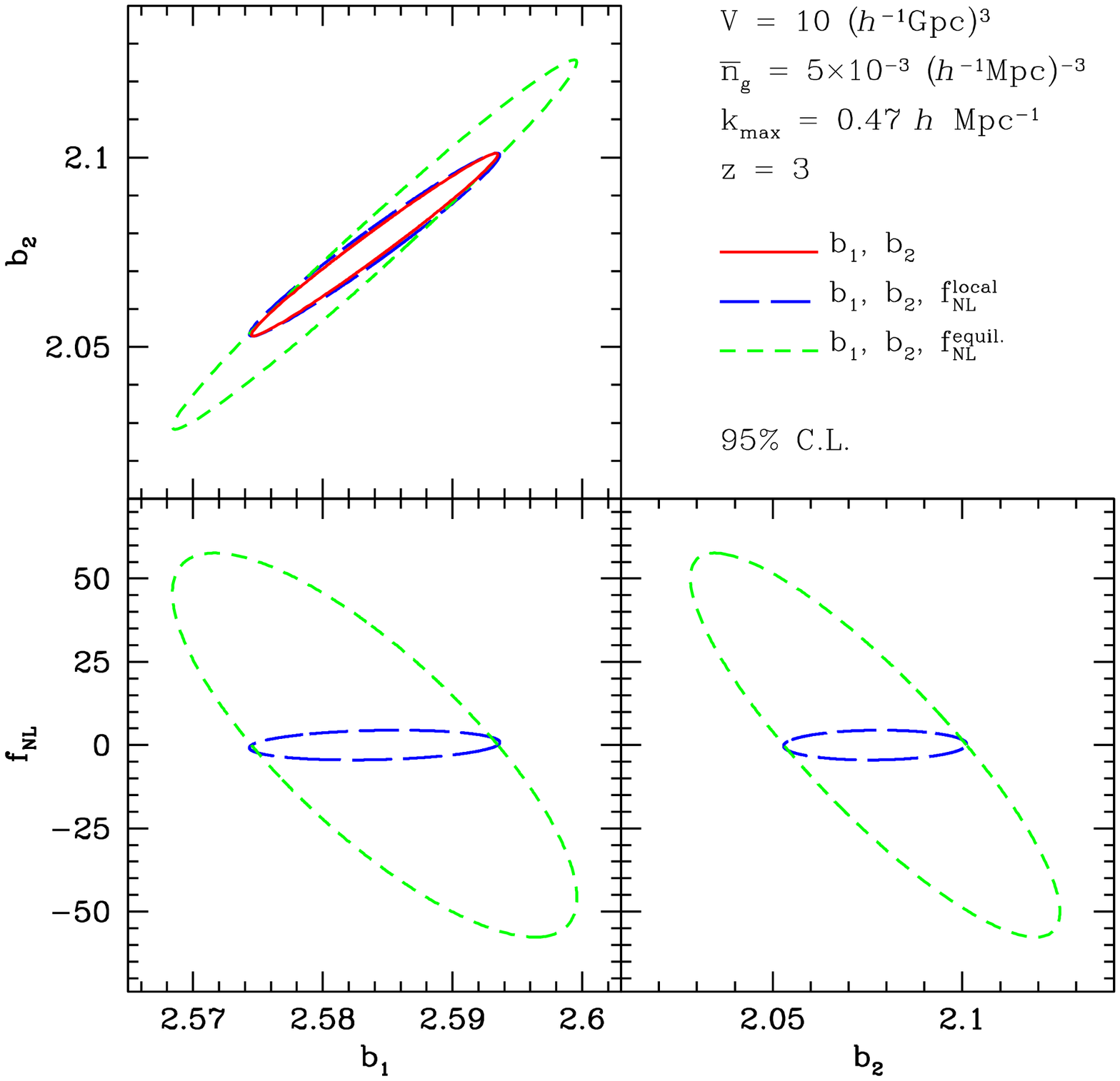}
\caption{\label{fig_contours} 
Two-dimensional joint 95\% C.L. constraints on galaxy bias and primordial 
non-Gaussianity from $z=1$ (left) and $z=3$ (right). The top left, bottom 
left, and bottom right panel show the joint constraints on ($b_1$, $b_2$), 
($b_1$, $\fNL$), and ($b_2$, $\fNL$), marginalized over $\fNL$, $b_2$, and 
$b_1$, respectively. The assumed survey volume is $V=10~\cGpc$, while the 
number density is $n_g=5\times 10^{-3}~\icMpc$. } 
\end{center}
\end{figure*}

Are galaxy bias and primordial non-Gaussianity independent? In Figure~
\ref{fig_contours} we show the 2-d joint constraints (95\% C.L.) on ($b_1$, 
$b_2$), ($b_1$, $\fNL$), and ($b_2$, $\fNL$), marginalized over $\fNL$, $b_2$, 
and $b_1$, respectively. The survey parameters are $V=10~\cGpc$, $n_g=5\times 
10^{-3}~\icMpc$, $z=1$ (left panels) and 3 (right panels). The fiducial values 
of bias parameters are $b_1=1.5$ and $b_2=0.035$ at $z=1$, and $b_1=2.6$ and 
$b_2=2.1$ at $z=3$.

We find that $\fNLl$ is not degenerate with $b_1$ or $b_2$, which is a very 
good news; however, $\fNLe$ reveals a rather strong degeneracy with both $b_1$ 
and $b_2$. Therefore, the equilateral model turns out to be much harder to 
constrain by CMB or galaxy surveys.

\subsection{Current and proposed redshift surveys}

We are now in a position to apply our Fisher matrix analysis tools to several 
current and future redshift galaxy surveys, both at low and high redshifts. 
For the sake of simplicity, and to allow for easier comparison, we shall 
assume that {\it each survey is characterized uniquely by its survey volume, 
$V$, redshift range, and galaxy number density, $n_g$}. In other words, we 
shall ignore complications related to the specific geometry and selection 
functions.

A galaxy survey of a large volume and a relatively low galaxy density (but not 
too low --- there should be at least $n_g\sim 1/P(\kMAX)$ galaxies for the 
optimal survey efficiency)  is necessary to detect the baryon acoustic 
oscillations (BAO) in the power spectrum, which may be used to constrain the 
nature of dark energy. We point out that such a survey should also provide 
competitive constraints on primordial non-Gaussianity. 
 
We shall consider the following survey designs. Table~\ref{tab_results_a} 
tabulates $V$, $n_g$, $z$, $\kMAX$, $b_1$ and $b_2$ of these surveys. We 
calculate $\kMAX$ from  $\sigma(R,z)= 0.5$ and $\kMAX=\pi/(2R)$ (Sec.~\ref{sec_fisher}), and $b_1$ and $b_2$ from a halo approach given in Sec.~
\ref{sec_bias}. 

\begin{itemize}
\item[(i)] Low $z$ surveys include two on-going surveys and one planned survey:
\begin{itemize}
\item The Sloan Digital Sky Survey (SDSS) main sample (for which a
      detailed analysis of the expected constraints on galaxy bias and
      primordial non-Gaussianity has been given in
      \citet{ScoccimarroSefusattiZaldarriaga2004}) at $\bar{z}=0$ and
      $V=0.3~\cGpc$.
\item The SDSS Luminous Red Galaxy (LRG) sample at $\bar{z}= 0.35$ and 
      $V=0.72~\cGpc$.
\item A proposed extension of SDSS, APO-LSS survey at $\bar{z}= 0.35$ and 
      $V=3.8~\cGpc$.
\end{itemize}

\item[(ii)] Intermediate $z$ surveys include three planned surveys:
\begin{itemize}
\item The Hobby-Eberly Dark Energy Experiment (HETDEX) \citep{HillEtal2004}, 
      targeting Lyman-$\alpha$ emitters (LAE) between $z=2$ and $z=4$. We 
      assume  $V=2.7 \cGpc$ and  $n_g=5\times 10^{-4}\icMpc$, following the 
      ``G2'' design given in \citet{TakadaKomatsuFutamase2006} for comparison.
\item Two redshift surveys with the planned Wide-Field Multi-Object
      Spectrograph (WFMOS),  one detecting 2 million galaxies at $0.5<z<1.3$ 
      in $2,000$ deg$^2$ (WFMOS1;  $V=4\cGpc$), and the other detecting 
      $600,000$ galaxies at $2.3<z<3.3$ in $300$ deg$^2$ (WFMOS2; $V=1\cGpc$) 
      \citep{GlazebrookEtal2005}. To facilitate comparison we  assume the same 
      number density of galaxies for both, $n_g=5\times 10^{-4}\icMpc$. 
\item The Advanced Dark Energy Physics Telescope (ADEPT) mission, a
      space-based redshift survey of 100 million galaxies at $1<z<2$ in
      $28,600$ deg$^2$. We  assume $V=100\cGpc$ and $n_g= 10^{-4}\icMpc$.
\end{itemize}

\item[(iii)] High $z$ surveys are represented by the Cosmic Inflation Probe 
      (CIP) mission,  a space-based redshift survey targeting H$\alpha$
      emitters at  $3.5<z<6.5$ \citep{MelnickEtal2004}. We assume $V=3.4\cGpc$ 
      and $n_g=5\times 10^{-3}\icMpc$, following the ``SG'' design given in 
      \citet{TakadaKomatsuFutamase2006}.
\end{itemize}

The 8th and 9th column in  Table~\ref{tab_results_a} tabulate $\Delta b_1$ and 
$\Delta b_2$ for Gaussian initial conditions ($\fNL=0$), the 10th, 11th, and 
12th column tabulate $\Delta b_1$, $\Delta b_2$, and $\Delta \fNLl$, and the 
13th, 14th, and 15th column tabulate $\Delta b_1$, $\Delta b_2$, and 
$\Delta \fNLe$. For HETDEX, WFMOS, ADEPT, and CIP we provide $\Delta b_1$, 
$\Delta b_2$, $\Delta \fNLl$, and $\Delta\fNLe$ from each redshift bin as well 
as $\Delta \fNLl$ and $\Delta \fNLe$ from a combined analysis of all bins. 

\begin{table*}[t]
\caption{\label{tab_results_a} 
Survey parameters ($V$ in units of $\cGpc$, $n_g$ in units of $\icMpc$, $z$), 
maximum wavenumber in the analysis ($\kMAX$ in units of $\kMpc$), fiducial 
values of galaxy bias ($b_1$ and $b_2$), and marginalized 1-$\sigma$ 
constraints from the Fisher matrix analysis of the reduced bispectrum in 
redshift space for  $\sigma_8=0.9$ and $n_s=1$.}
\begin{ruledtabular}
\begin{tabular}{l|ccc|c|cc||cc||ccc||ccc}
        & $V$  &$n_g$& $z$  &$\kMAX$&$b_1$ &$b_2$  & $\dbo$& $\dbt$& $\dbo$& $\dbt$&$\dfNLl$&$\dbo$ & $\dbt$&$\dfNLe$\\
\hline\hline
SDSS    &$0.3$ &$30$ &$0$   & $0.09$&$1.19$&$-0.10$&$0.270$&$0.151$&$0.309$&$0.151$&$ 255.5$&$0.450$&$0.421$&$1775$  \\
\hline
LRG     &$0.72$& $1$ &$0.35$& $0.11$&$2.14$&$ 0.96$&$0.209$&$0.348$&$0.223$&$0.353$&$ 113.4$&$0.338$&$0.726$&$ 998$  \\
\hline
APO-LSS &$3.8 $& $4$ &$0.35$& $0.11$&$1.69$&$ 0.21$&$0.069$&$0.068$&$0.071$&$0.069$&$  34.9$&$0.108$&$0.160$&$ 386$  \\
\hline
WFMOS1  &$1.6$ & $5$ &$0.7$ & $0.14$&$1.87$&$ 0.45$&$0.076$&$0.096$&$0.080$&$0.096$&$  41.0$&$0.123$&$0.216$&$ 435$  \\
        &$2.4$ & $5$ &$1.1$ & $0.18$&$2.16$&$ 1.00$&$0.047$&$0.081$&$0.048$&$0.081$&$  23.1$&$0.076$&$0.175$&$ 266$  \\
        &\multicolumn{3}{l}{combined} & &  &       & -     & -     & -     & -     &$  20.1$& -     & -     &$ 227$  \\
\hline
ADEPT   &$45$  & $1$ &$1.25$& $0.20$&$2.97$&$ 3.44$&$0.020$&$0.063$&$0.021$&$0.063$&$   6.1$&$0.031$&$0.111$&$  73$  \\
        &$55$  & $1$ &$1.75$& $0.26$&$3.44$&$ 5.43$&$0.017$&$0.066$&$0.017$&$0.067$&$   4.5$&$0.025$&$0.112$&$  53$  \\ 
        &\multicolumn{3}{l}{combined} & &  &       & -     & -     & -     & -     &$   3.6$& -     & -     &$  43$  \\
\hline
WFMOS2  &$0.5$ & $5$ &$2.55$& $0.38$&$3.27$&$ 4.64$&$0.058$&$0.220$&$0.060$&$0.223$&$  25.7$&$0.094$&$0.406$&$ 256$  \\ 
        &$0.5$ & $5$ &$3.05$& $0.48$&$3.64$&$ 6.39$&$0.056$&$0.253$&$0.058$&$0.255$&$  22.1$&$0.087$&$0.439$&$ 215$  \\
        &\multicolumn{3}{l}{combined} & &  &       & -     & -     & -     & -     &$  16.8$&$   $  & -     &$ 164$  \\
\hline
HETDEX  &$0.68$& $5$ &$2.25$& $0.34$&$3.05$&$ 3.70$&$0.051$&$0.172$&$0.053$&$0.174$&$  23.6$&$0.083$&$0.326$&$ 244$  \\
        &$0.69$& $5$ &$2.75$& $0.42$&$3.42$&$ 5.32$&$0.049$&$0.199$&$0.050$&$0.201$&$  20.0$&$0.077$&$0.357$&$ 202$  \\
        &$0.67$& $5$ &$3.25$& $0.53$&$3.79$&$ 7.16$&$0.050$&$0.237$&$0.051$&$0.238$&$  18.0$&$0.076$&$0.401$&$ 177$  \\
        &$0.64$& $5$ &$3.75$& $0.65$&$4.14$&$ 9.20$&$0.053$&$0.291$&$0.054$&$0.292$&$  17.1$&$0.079$&$0.469$&$ 163$  \\
        &\multicolumn{3}{l}{combined}& & - & -     & -     & -     & -     & -     &$   9.6$& -     & -     &$  95$  \\
\hline
CIP     &$1.26$&$50$ &$4$   & $0.71$&$3.16$&$ 4.12$&$0.010$&$0.036$&$0.010$&$0.037$&$   4.7$&$0.016$&$0.066$&$  51$  \\
        &$1.13$&$50$ &$5$   & $1.03$&$3.72$&$ 6.76$&$0.010$&$0.047$&$0.010$&$0.048$&$   4.0$&$0.015$&$0.079$&$  40$  \\
        &$1.02$&$50$ &$6$   & $1.46$&$4.26$&$ 9.90$&$0.011$&$0.066$&$0.012$&$0.066$&$   3.8$&$0.016$&$0.102$&$  36$  \\
        &\multicolumn{3}{l}{combined}& & - & -     & -     & -     & -     & -     &$   2.4$& -     & -     &$  24$  \\
\end{tabular}
\end{ruledtabular}
\end{table*}

The predicted constraints on galaxy bias range from a few percent for APO-LSS, 
WFMOS and HETDEX to less than $1\%$ for ADEPT, the latter being significantly 
better owing obviously to a larger survey volume. As we have mentioned already 
in Sec.~\ref{sec:deg}, inclusion of $\fNLl$ does not degrade sensitivity to 
galaxy bias, whereas $\fNLe$ does degrade it significantly.

Our prediction for SDSS should not be compared directly to those derived by 
\citet{ScoccimarroSefusattiZaldarriaga2004}, as we have  ignored covariance 
between different bispectrum configurations due to non-linear effects and 
survey geometry (Sec.~\ref{sec_covar}), which was included in their work. 
In this sense our analysis is more optimistic; however, in the other sense our 
analysis is in fact more realistic than theirs. We have used 
$\kMAX\simeq 0.09\kMpc$ for SDSS, which is significantly more conservative 
than their value, $\kMAX = 0.3\kMpc$. This has made our predicted errors much 
weaker than theirs: they obtained $\Delta b_1/b_1=0.04$ (ours $0.26$) and 
$\dfNLl=145$ (ours $256$). 

How would these results depend on $\sigma_8$ and $n_s$? A lower value for the 
rms density fluctuations, $\sigma_8=0.75$, as recently suggested by the WMAP 
3yr data \citep{SpergelEtal2006}, increases $\kMAX$ because non-linearity is 
weaker for a lower $\sigma_8$, and thus improves the constraints. On the other 
hand, it follows from equation (\ref{stn_b1}) that the signal-to-noise for a 
given triangle configuration is proportional to $\sigma_8^2$, and thus a lower 
$\sigma_8$ results in worse constraints. When combined, these two effects 
result in 20\% and 30\% stronger constraints on galaxy bias and primordial 
non-Gaussianity, respectively, for low-redshift surveys, whereas these effects 
cancel for intermediate and high $z$ surveys. A departure from a scale 
invariant spectrum has  a smaller effect. For $n_s=0.95$ we find only  $10\%$ 
improvement in bias and non-Gaussianity.

\section{Constraining the HOD}
\label{sec_hod}

So far we have assumed that the linear and quadratic bias, $b_1$ and $b_2$, 
are completely free. For surveys spanning a large redshift range, 
therefore, we had to introduce multiple redshift bins, and, as a result, we 
had to have an excessive number of free parameters, two times the number of 
redshift bins. 

In this section we attempt to reduce the number of free parameters by using
a halo approach. The fiducial values of $b_1$ and $b_2$ were derived from a 
given form of the HOD (Sec.~\ref{sec_bias}). If we can parametrize the HOD by 
fewer parameters than two times the number of redshift bins, the model has 
more constraining power.

In the limit that the evolution of bias is given by the mass function, we may 
approximate that the HOD is independent of $z$. We make a minimal extension of 
the HOD that we used in Sec.~\ref{sec_bias}; namely, instead of fixing a ratio 
of $M_1$ and $M_{min}$, we let it free:
\beq
\log_{10}M_1=a+\log_{10}M_{min}.
\eeq
The fiducial value is $a=1.1$, as before. We still assume that a relation 
between $M_{cut}$ and $M_1$ is given by equation (\ref{Mcut}). We then replace 
$b_1$ and $b_2$ at different redshift bins with a single parameter $a$ in the 
Fisher matrix analysis. 

The expected 1-$\sigma$ errors on $a$ and primordial non-Gaussian parameters 
are given in Table~\ref{Tab_hod} in the third to fifth columns. We find that 
the error on $\fNLl$ is  unaffected: this is an expected result because 
$\fNLl$ is not degenerate with galaxy bias. On the other hand, we find a 
significant, about a factor of two, improvement in $\fNLe$. This is due to the 
fact that the analysis in terms of the HOD is equivalent to introducing a 
theoretical prior on a relation between $b_1$ and $b_2$, lifting the 
degeneracy. In other words, while in the previous section we allowed $b_1$ and 
$b_2$ to vary independently, we are now making use of the fact that the Halo 
Model predicts them to be strongly correlated.

\begin{table*}[t]
\squeezetable
\caption{\label{Tab_hod} Marginalized 1-$\sigma$ errors on the HOD parameters 
and primordial non-Gaussianity from the Fisher matrix analysis of the reduced 
bispectrum in redshift space for $\sigma_8=0.9$ and $n_s=1$. The survey 
parameters and $\kMAX$ are the same as in Table~\ref{tab_results_a}.}
\begin{ruledtabular}
\begin{tabular}{l|c||c|cc|cc||cc|ccc|ccc}
\multicolumn{2}{l}{Survey} & \multicolumn{5}{c}{1-parameter HOD} & \multicolumn{8}{c}{2-parameter HOD} \\
\hline
        &$z$    &$\De a$ &$\De a$ &$\dfNLl$&$\De a$ &$\dfNLe$&$\De a$&$\De b$ &$\De a$&$\De b$&$\dfNLl$&$\De a$&$\De b$&$\dfNLe$\\
\hline
SDSS    &$0$    &$0.20$  &$0.26$  &$198$   &$0.20$  & $ 500$ & -     & -      & -     & -      & -      & -     & -      & -    \\
\hline
LRG     &$0$    &$0.10$  &$0.15$  &$112$   &$0.13$  & $ 363$ & -     & -      & -     & -      & -      & -     & -      & -    \\
\hline
APO-LSS &$0.25$ &$0.064$ &$0.077$ &$35.1$  &$0.066$ & $ 120$ & -     & -      & -     & -      & -      & -     & -      & -    \\
\hline
WFMOS1  &$0.7$  &$0.050$ &$0.066$ &$39.0$  &$0.057$ & $ 134$ &$ 540$ &$   780$&$  570$&$   820$&$  41$  &$ 1800$&$ 2500$ &$435$ \\
        &$1.1$  &$0.024$ &$0.031$ &$22.7$  &$0.029$ & $  91$ &$1050$ &$   960$&$ 1070$&$   970$&$  23$  &$ 3100$&$ 2800$ &$266$ \\
        & comb. &$0.024$ &$0.028$ &$19.6$  &$0.026$ & $  75$ &$ 0.14$&$  0.14$&$ 0.15$&$  0.14$&$  20$  &$ 0.15$&$  0.14$&$ 75$ \\
\hline
ADEPT   &$0.7$  &$0.0050$&$0.0062$&$ 6.1$  &$0.0078$& $  35$ &$  56$ &$   45$ &$   56$&$    45$&$ 6.1$  &$  116$&$   93$&$  73$ \\
        &$1.1$  &$0.0033$&$0.0040$&$ 4.5$  &$0.0050$& $  28$ &$ 1200$&$   710$&$ 1200$&$   710$&$ 4.5$  &$ 2400$&$ 1300$&$  53$ \\
        & comb. &$0.0033$&$0.0034$&$ 3.6$  &$0.0042$& $  22$ &$0.019$&$ 0.012$&$0.020$&$ 0.012$&$ 3.6$  &$0.020$&$0.012$&$  22$ \\
\hline
WFMOS2  &$0.7$  &$0.0147$&$0.020$ &$26$    &$0.021$ & $ 120$ &$   83$&$    33$&$  83$ &$    33$&$  26$  &$  177$&$   70$&$ 256$ \\
        &$1.1$  &$0.0120$&$0.016$ &$22$    &$0.017$ & $ 110$ &$  789$&$   259$&$ 789$ &$   259$&$  22$  &$ 1539$&$  505$&$ 215$ \\
        & comb. &$0.0120$&$0.013$ &$17$    &$0.013$ & $  81$ &$0.11$ &$ 0.038$&$ 0.11$&$ 0.038$&$  17$  &$0.11$ &$0.038$&$  81$ \\
\hline
HETDEX &$2.25$ &$0.015$ &$0.020$ &$24.1$  &$0.021$ & $111$  &$3500$ &$1600$  &$3500$ &$1600$  &$ 24.1$ &$7900$ &$3500$  &$ 250$\\
        &$2.75$ &$0.011$ &$0.015$ &$19.9$  &$0.016$ & $ 99$  &$1400$ &$ 500$  &$1400$ &$ 500$  &$ 19.9$ &$2800$ &$1000$  &$ 201$\\
        &$3.25$ &$0.010$ &$0.013$ &$18.0$  &$0.014$ & $ 93$  &$3200$ &$ 990$  &$3200$ &$ 990$  &$ 18.0$ &$6100$ &$1900$  &$ 177$\\
        &$3.75$ &$0.010$ &$0.013$ &$17.2$  &$0.013$ & $ 92$  &$ 730$ &$ 200$  &$ 730$ &$ 200$  &$ 17.2$ &$1300$ &$ 350$  &$ 164$\\
        & comb. &$0.005$ &$0.007$ &$ 9.7$  &$0.008$ & $ 49$  &$0.034$&$0.11$  &$0.034$&$0.011$ &$  9.7$ &$0.035$&$0.011$ &$  49$\\
\hline
CIP     & $4$   &$0.0032$&$0.0041$&$ 4.7$  &$0.0042$& $ 23$  &$ 11.1$&$ 2.8$  &$11.1$ &$2.8$   &$   4.7$&$24$   &$6.0$   &$  51$\\
        & $5$   &$0.0026$&$0.0034$&$ 4.0$  &$0.0034$& $ 21$  &$  6.9$&$ 1.4$  &$ 6.9$ &$1.4$   &$   4.0$&$13$   &$2.6$   &$  40$\\
        & $6$   &$0.0025$&$0.0033$&$ 3.7$  &$0.0034$& $ 21$  &$ 13.1$&$ 2.2$  &$13.1$ &$2.2$   &$   3.8$&$22$   &$3.7$   &$  36$\\
        & comb. &$0.0016$&$0.0021$&$ 2.3$  &$0.0021$& $ 13$  &$0.010$&$0.0020$&$0.011$&$0.0020$&$   2.4$&$0.011$&$0.0020$&$  13$\\
\end{tabular}
\end{ruledtabular}
\end{table*}

When we have a survey that covers a wide range in $z$, we may be able to
constrain more-than-one parameters in the HOD. To see how it works we extend 
the minimal model by introducing one more parameter:
\beq
\log_{10}M_1=a+b z+\log_{10}M_{min},
\eeq
and we assume $b=0$ as the fiducial value. An analysis based on a single 
redshift bin would naturally lead to a large degeneracy between $a$ and $b$. 
However, one can lift this degeneracy by including multiple redshift bins.

In Table~\ref{Tab_hod}, columns 6th to 13th, we present the expected 
1-$\sigma$ errors on $a$ and $b$ as well as on primordial non-Gaussianity. One 
can clearly see that $a$ and $b$ are degenerate within a single redshift bin: 
combined errors are orders of magnitude smaller than those from single bins. 
The determination of primordial non-Gaussianity is largely unaffected by the 
extra HOD parameter.

\section{Conclusions}
\label{sec_conclusions}

The quest to understand the nature of dark energy has recently provided 
a further motivation for future large redshift surveys. It is certain that
the study of higher order correlation functions of galaxies will be 
required in order to extract maximum cosmological information from 
such large data sets. For instance, as it has been recognized for more than a 
decade, the bispectrum can be used to measure non-linearities in the 
galaxy-mass relation. Non-linearity in galaxy bias must be understood better to
meet  the high accuracy required for precision measurements of the baryon 
acoustic oscillations and their interpretation as a standard ruler, 
particularly for highly biased tracers such as the luminous red galaxies
and Lyman-$\alpha$ emitters.

We have shown that, with the most conservative assumption about the
maximum wavenumber, $\kMAX$, used in the analysis, the bispectrum measured 
from a galaxy redshift survey should yield a fractional error on the linear 
bias of order $0.1\sqrt{\cGpc/V}$ at $z=0$ to $0.05\sqrt{\cGpc/V}$ at $z=6$.
This is an {\it extremely} conservative limit, however, 
as it assumes no understanding of even the mildest non-linearities in the
dark matter and galaxy biasing evolution. 

An analysis that includes all configurations down to mildly non-linear
scales, 
within  which $\sigma(R)\lesssim 0.5$ is satisfied, should yield more than an
order of magnitude better determination of bias parameters, both linear
and non-linear, owing to  the large number of configurations 
available at smaller scales. This is precisely where intermediate to
high-$z$ galaxy surveys play a leading role: non-linearity is much
weaker at higher $z$, and therefore we can access to a large number of
modes on small scales.

While we can study non-linearities in the matter 
distribution by means of new techniques based on perturbation theory and 
of N-body simulations, a simple, local description of 
galaxy bias may have to be improved further. 
In this perspective, our results show that a large amount of information
can be extracted from higher-order correlations which, in turn,
may be used to constrain more sophisticated models of galaxy biasing.  

We have also shown that the bispectrum from large-volume,
high-redshift surveys is highly sensitive to primordial non-Gaussianity. 
The CMB observations have been the best probe of the Gaussian nature
of primordial perturbations so far, and the 
Planck satellite would be quite close 
to the ideal experiment limit. 
On the other hand, a redshift survey of the large scale structure
 actually contains much more
information than CMB, as the number of modes available from 
the 
three-dimensional fluctuations is vastly larger than that from the
two-dimensional temperature and polarization anisotropies.

Not only can they provide independent constraints on scales 
smaller than those probed by CMB, but also their constraints can be 
comparable to, if not better than, those from CMB.
The best limit one can achieve from  
an all-sky survey up to redshift $\sim 5$ should reach $\fNLl\sim 0.2$ and 
$\fNLe\sim 2$, an order of magnitude better than the best limits achievable 
by CMB. The planned surveys such as HETDEX and ADEPT should reach the 
constraints that are comparable to those from the WMAP and Planck CMB 
experiments, respectively. It should also be understood that galaxy surveys 
provide the best limits on small scales that are not accessible by CMB.
This is particularly important when
probing scale-dependent non-Gaussian models. 

Finally, we have shown that galaxy bias parameters modeled by the halo 
occupation distribution should be very useful for lifting 
parameter degeneracies between non-linear gravity, galaxy bias,
and primordial non-Gaussianity.

We believe that, while a lot of work still has to be done in order 
to achieve a satisfactory description of the evolution of non-linearities and 
galaxy bias at small scales, higher order correlation functions will play a
crucial role in the analysis of future redshift surveys, providing 
indispensable information on galaxy bias as well as on the nature of 
primordial perturbations.

\begin{acknowledgments}
E.S. would like to thank the Department of Astronomy of the University of
Texas at Austin for the very kind hospitality, Rom\'an Scoccimarro and Josh 
Frieman for useful comments on an earlier draft of the paper and Charlie 
Conroy for discussions. E.S. is supported by the U.S. Department of Energy
and by NASA grant NAG 5-10842 at Fermilab. E.K. acknowledges support from the 
Alfred P. Sloan Foundation.
\end{acknowledgments}

\bibliography{BispHighZ}

\end{document}